\documentclass[]{aa}  
\usepackage{graphicx}
\usepackage{txfonts}
\usepackage{hyperref}
\usepackage{xcolor}

\begin{document}

   \title{Tidal distortion and disruption of rubble-pile bodies revisited}

   \subtitle{Soft-sphere discrete element analyses}

   \author{Yun Zhang\inst{1}
          \and
          Patrick Michel\inst{1}
          }

   \institute{$^1$ Universit\'e C\^ote d'Azur, Observatoire de la C\^ote d'Azur, CNRS, Laboratoire Lagrange, Nice, France.
              \email{\href{mailto:yun.zhang@oca.eu}{yun.zhang@oca.eu}}
             }

   \date{Received 29 February 2020 / Accepted for publication in A\&A}

 
  \abstract
   {In the course of a  close approach to planets or stars, the morphological and dynamical properties of rubble-pile small bodies can be dramatically modified, and some may catastrophically break up,  as in the case of comet Shoemaker-Levy 9.  This phenomenon is of particular interest for the understanding of the evolution and population of small bodies, and for making predictions regarding the outcomes of future encounters.  Previous numerical explorations have typically  used methods that do not adequately represent the nature of rubble piles.  The encounter outcomes and influence factors are still poorly constrained.}
   {Based on recent advances in modeling rubble-pile physics, we aim to provide a better understanding of the tidal encounter processes of rubble piles through soft-sphere discrete element modeling (SSDEM) and to establish a database of encounter outcomes and their dependencies on encounter conditions and rubble-pile properties.}
   {We performed thousands of numerical simulations using the SSDEM implemented in the $N$-body code {\it pkdgrav}  to study the dynamical evolution of rubble piles during close encounters with the Earth. The effects of encounter conditions, material strength, arrangement, and resolution of constituent particles are explored.}
   {Three typical tidal encounter outcomes are classified, namely: deformation, mass shedding, and disruption, ranging from mild modifications to severe damages of the progenitor. The outcome is highly dependent on the encounter conditions and on the structure and strength of the involved rubble pile. The encounter speed and distance required for causing disruption events are much smaller than those predicted by previous studies, indicating a smaller creation rate of tidally disrupted small body populations. Extremely elongated fragments with axis ratios $\sim$1:6 can be formed by moderate tidal encounters. Our analyses of the spin-shape evolution of the largest remnants reveal reshaping mechanisms of rubble piles in response to tidal forces, which is consistent with stable rubble-pile configurations derived by continuum theory. A case study for Shoemaker-Levy 9 suggests a low bulk density (0.2--0.3 g/cc) for its progenitor.}
   {}


   \maketitle
%
\section{Introduction}
\label{s:introduction}

 The dynamical evolution of small bodies in planetary systems can sometimes lead to very close approaches to planets or stars that can either cause surface motion of the involved small body, modify its shape, or even catastrophically disrupt it. In fact, tidal disruption events do not only occur in our Solar System, but they are also common in exoplanetary systems and in different stages of star evolution \citep{Jura2003, Gillon2014}. The outcomes of tidal encounters require a good understanding as they may be at the origin of some of the observed characteristics of small bodies and, therefore, these may give important clues on their history and structures \citep{Bottke1999, Binzel10, Zhang2020}. In this paper, we revisit this process with a more realistic numerical treatment than in previous works.

 The most striking evidence of tidal disruption was provided by the encounter of comet Shoemaker-Levy 9 (SL9) with Jupiter that broke the body into 21 pieces when it approached within 1.33 Jovian radii of the planet center in 1992 \citep{Sekanina1994}. These pieces eventually collided with the giant planet during the subsequent encounter in July 1994.  

 Some asteroids belonging to the population of near-Earth objects (NEOs)  also experience very close encounters with the Earth during their evolution in the near-Earth space.  A famous example in the near future is the predicted approach of Apophis at $5.7 \pm 1.4$ Earth radii from our planet in April 13, 2029, which should make this 340-meter-diameter asteroid visible with the naked eye from the ground \citep{Brozovic2018}. The possible tidal effects during this passage are still under debate and may just be very local on the asteroid's surface \citep{Yu2014}, but they may be measurable thanks to an in-situ seismometer \citep{DeMartini2019}.

 Close approaches of NEOs and other bodies to planets or stars can have a large range of possible configurations. Possible tidal effects and their outcomes highly depend on the initial conditions of those encounters as well as on the internal structure and other characteristics of the encountered bodies. For a fluid, the Roche limit \citep{Roche1847} is a frequently used, well-known concept, which defines the closest approach for a small body to a large one within which the smaller body cannot withstand the tidal forces and must, therefore, break up. Using a static theory, \cite{Holsapple2006,Holsapple2008} predicted the change in shape and possible failure of small bodies, modeled as continuum solid bodies with or without cohesion, as a function of the encounter distance to a planet, aspect ratios of the small body's shape, its angle of friction, and level of cohesion. This static theory has led to an extensive database of conditions for the onset of disruption. However, the nature of the resulting dynamics of a body as it disintegrates, that is, how many fragments are generated, what  their resulting shapes are and how they eventually evolve individually, or how they reaccumulate, could not be assessed and still requires the explicit modeling of the dynamics of the encounter.

 Several pieces of evidence point to a rubble-pile structure for most small bodies \citep{Richardson2002}, which was later confirmed by a number of space missions \citep[e.g.,][]{Fujiwara2006, Lauretta2019, Watanabe2019}. In fact, due to their small size (smaller than a few tens of km), most of them are expected to be formed from the collisional disruption of a parent body in the main belt, leading to their formation as aggregates as a result of gravitational reaccumulation of smaller pieces resulting from the disruption \citep{Michel2001, Schwartz2018}. Once formed, these bodies can be injected in dynamical resonances that transport them to the near-Earth space and become NEOs \citep{Gladman1997,Granvik2018}. 

 The numerical modeling of the dynamics of close Earth approaches of rubble piles, modeled as gravitational aggregates, was done by \cite{Richardson1998}, who reviewed past work at that time and explored a large parameter space, including the NEO’s periapse distance, encounter velocity with the Earth, initial spin-axis orientation, and body orientation at periapse. The progenitor was modeled as an aggregate of 247 identical rigid spherical particles, each 255 m in diameter. The number of particles was chosen as a compromise between resolution and computational efficiency at the time. A gravitational $N$-body code developed by \cite{Richardson1993, Richardson1994, Richardson1995} was used to integrate the equations of motion. The collisions between the particles were treated as hard-sphere collisions, with no friction \citep[see][for details]{Richardson1998}. Their model, which they recognized to be crude, provided a first general understanding of tidal encounter processes of rubble-pile objects. 

 Further numerical modeling of tidal encounters was then performed by \cite{Walsh2006}, whose motivation was to determine whether such encounters could explain the formation of observed binary asteroids. Covering a range of encounter conditions, they performed with the gravitational parallel $N$-body tree code {\it pkdgrav} 110,500 simulations, using a rubble-pile progenitor consisting of $\sim$1000 rigid 150-m spheres bound to one another by gravity. They then gave to those progenitors one of four rotation rates, that is, 3-, 4-, 6-, or 12-h periods. Their results show distributions of physical and orbital properties of the resulting binaries, which were compared to the observed ones at that time, showing that this process may not be the source of most binaries. The main difference with observed binaries was that the eccentricity of the secondary formed in their simulations is much higher than the one of actual binaries. The formation of small binaries was then found to originate from another mechanism, which is beyond the scope of the current paper \citep[see][for a review]{Walsh2015}.

 To understand the properties and evolution of NEO families originated by tidal fragmentation, tidal disruption processes and outcomes were also numerically investigated by \cite{Schunova2014}. Using the same code {\it pkdgrav}, they carried out $N$-body simulations of the tidal disruptions of km-sized spherical rubble-pile progenitors on a range of Earth-crossing orbits. Rubble-pile models consisting of above 2000 solely gravitationally-bound identical rigid 60-m-radius spheres with densities of 3.34 g/cc were used in their study. Their simulation results show that tidal disruption processes are capable of creating detectable NEO families. However, due to the chaotic dynamical evolution of small bodies in the near-Earth space, these NEO families can only be identified by their orbital similarity within several tens of thousands of years. According to the null detection of NEO families \citep{Schunova2012}, they derived a lower limit on the creation rate of tide-producing NEO families, that is, the frequency for a km-sized NEO to be tidally disrupted and create a family is less than once per $\sim$2300 years.

 These past works have revealed the significance of tidal disruption processes with regard to the common understanding of the small body population and evolution. However, they were all performed using a method called the hard-sphere discrete element method (HSDEM) to treat collisions between the spherical particles constituting the rubble piles \citep[see][]{Richardson2011}. More recently, because of an increased interest to understand the dynamics of regolith on small body surfaces, a new method developed for granular material studies has been used in the field, called the soft-sphere discrete element method (SSDEM), which solves for all contact forces between the particles constituting a granular medium, accounting for the various kinds of frictions between them \citep[see][for a review]{Murdoch2015}. In the SSDEM, macroscopic particles are treated as deformable spheres, allowing overlaps between particles to act as proxies for actual deformation. Particles are taken to be in contact if and only if their surfaces are touching or mutually penetrating. The greater the extent of this penetration, the more repulsive the force that is generated. All contact forces are then computed, accounting for a static/dynamic tangential, as well as rolling and twisting frictions. \cite{Schwartz2012} implemented the SSDEM in the code {\it pkdgrav}. \cite{Zhang2017} updated this implementation by introducing the shape parameter $\beta$ to represent a statistical measure of the contact area of realistic particle shapes, which allows mimicking the strong rotational resistance between irregular particles.  

 In this study, we use the SSDEM to model tidal encounters, and investigate the influence of various encounter conditions and simulation parameters. In particular, as the material strength can significantly affect the mechanical properties and dynamical behaviors of self-gravitating rubble piles \citep{Zhang2018}, we investigate the effects of the material shear strength of simulated rubble piles. Furthermore, the packing commonly used to model the aggregates in previous studies \citep[e.g.,][]{Richardson1998, Walsh2008, Schunova2014, DeMartini2019} is the hexagonal close packing (HCP), while it is likely that the internal configuration of a rubble pile formed by reaccumulation is closer to the one corresponding to a random close packing \cite[RCP; e.g., see][]{Michel2013}. So, we investigate the effect of particle configuration by performing the same campaign of simulations using both RCP and HCP models. We also investigate the resolution effect by considering aggregates composed of a different number of particles.  We then compare our results to the continuum theory as well as to past HSDEM simulation findings. As we will show, the outcomes are very different from those obtained by using the HSDEM.  We also apply our SSDEM modeling to the tidal encounter of SL9 with Jupiter  and compare our findings with those obtained by a polyhedral rubble-pile approach \citep{Movshovitz2012}.

 Section \ref{s:method} presents our methodology and simulation setups, and Sect.~\ref{s:results} exposes our simulation results using the SSDEM. The comparison with past work using the HSDEM is presented in Sect.~\ref{s:ComparingHS}. A case application for SL9 is shown in Sect.~\ref{s:sl9}. Section \ref{s:conclusions} gives our conclusions and perspectives. 

\section{Methodology}
\label{s:method}

 We use the parallel $N$-body tree code {\it pkdgrav} and its implementation of the soft-sphere  discrete element method (SSDEM). We refer to \cite{Schwartz2012} and \cite{Zhang2017} for details on this implementation. Here is is worth noting that a granular physics model, including four dissipation and friction components in the normal, tangential, rolling, and twisting directions, is applied for computing particle contact forces. The compressive strength of the material is controlled by two stiffness constants, ($k_N$, $k_S$), for the normal and tangential directions; the contact energy dissipation is controlled by two coefficients of restitution, ($\varepsilon_N$, $\varepsilon_S$), for the normal and tangential directions; the magnitude of material shear strength is controlled by three friction coefficients for the tangential, rolling, and twisting directions, ($\mu_S$, $\mu_R$, $\mu_T$), along with a shape parameter that represents a statistical measure of real particle shapes, $\beta$. 

\subsection{Rubble-pile models}
\label{s:method:rubblepile}

 The rubble-pile progenitor is modeled as a 1-km-radius spherical granular assembly consisting of $\sim$10,000 identical spherical particles. These constituent particles interact through long-range gravitational forces and short-range physical contact forces \citep{Zhang2017}. Regarding the internal packing, we explore two possible configurations in this study: an HCP configuration (with $N=11$,577 and a particle radius of 40 m) and an RCP configuration (with $N=10$,011 and a particle radius of 40 m). In order to make direct comparisons with the HSDEM results, the initial conditions of the simulated rubble piles are set to the same as those used in \cite{Schunova2014}. The particle densities for the HCP and RCP models are set to 3.34 g/cc and 3.85 g/cc, respectively, which yield an alike internal bulk density of $\sim$2.47 g/cc as the interior packing efficiency is $\sim$0.739 and $\sim$0.641 for the HCP and RCP models, respectively. The initial mass $M_0\sim10^{13}$ kg. Prior to the encounter, the spherical progenitor has a prograde rotation with a spin period of $P_0=4.3$ h. We also perform a series of simulations using an HCP model with a lower number of particles (with $N=2$,953 and a particle radius of 60 m) to investigate the influence of the resolution on the outcomes. This low-resolution HCP model is almost identical to the rubble-pile model used in \cite{Schunova2014}, allowing us to make a one-to-one comparison of our SSDEM and their HSDEM results (see Sect. \ref{s:ComparingHS}).

\subsection{Material parameters}
\label{s:method:material}

 Our SSDEM contact model as well as the relation between model parameters and material strength have been calibrated with laboratory experiments on real sands \citep{Schwartz2012, Zhang2018}. The normal stiffness $k_N$ is set to $3.9\times10^9$ N/m, equivalent to a Young's modulus of $\sim$30 MPa \citep{DeMartini2019}. This value is within the measured strengths of meteorites \citep{Pohl2020}. The tangential stiffness $k_S$ is set to $(2/7)k_N$ to keep normal and tangential oscillation frequencies equal to each other. To precisely integrate the interactions between particles, the timestep $\Delta t$ is set to 0.03 s for the simulations considered in this study. The coefficients of restitution,  $\varepsilon_N$ and $\varepsilon_S$, are set to 0.55, resembling the energy dissipation behavior of terrestrial rocks \citep{Chau2002}.

 Previous analytical studies show that the structural stability of granular small bodies during a tidal encounter inherently depends on the angle of friction, which represents the slope that specifies shear strength under a given pressure \citep[e.g.,][]{Dobrovolskis1990, Holsapple2006, Sharma2006}. In our SSDEM model, there are four parameters, ($\mu_S$, $\mu_R$, $\mu_T$, $\beta$), that give rise to shear strength. As suggested by \cite{Jiang2015}, the static friction coefficients for rolling and twisting, $\mu_R$ and $\mu_T$, are set to 1.05 and 1.3, respectively, corresponding to sand particles of medium hardness. The two free parameters, that is, the tangential friction coefficient, $\mu_S$, and the shape parameter, $\beta$, are  used to adjust the material friction angle $\phi$ \citep{Zhang2017}.

 For the SSDEM HCP model, the considered set of ($\mu_S$, $\beta$) is (0.2, 0.3), which corresponds to a material having a low surface friction. However, a high friction angle ($\phi\sim40^\circ$) is measured for the SSDEM HCP model because of the interlocking geometry of its components \citep{Zhang2017}. For the HSDEM HCP model, because contact forces are not physically computed, the angle of friction is also controlled by the interlocking geometry and therefore its angle of friction $\phi\sim40^\circ$ \citep[see, e.g.,][]{Richardson2005}, which is similar to the material represented by our SSDEM HCP model. This allows us to make a one-to-one comparison of our SSDEM HCP model behaviors and past studies using the HSDEM HCP model (see Sect.~\ref{s:ComparingHS}).

 For the RCP model, which is a more natural representation of a granular small body's internal configuration, the friction angle is mainly controlled by the material friction parameters \citep{Zhang2017}. We consider the three sets of ($\mu_S$, $\beta$) to explore the effect of material shear strength, where (0.2, 0.3) corresponds to a friction angle of $\phi=18^\circ$, (0.6, 0.5) corresponds to a friction angle of $\phi=27^\circ$, and (1.0, 0.8) corresponds to a friction angle of $\phi=32^\circ$.

\subsection{Tidal encounter setup}

\label{s:method:tidalencounter}
   \begin{figure}
    \centering
    \includegraphics[width = 8 cm]{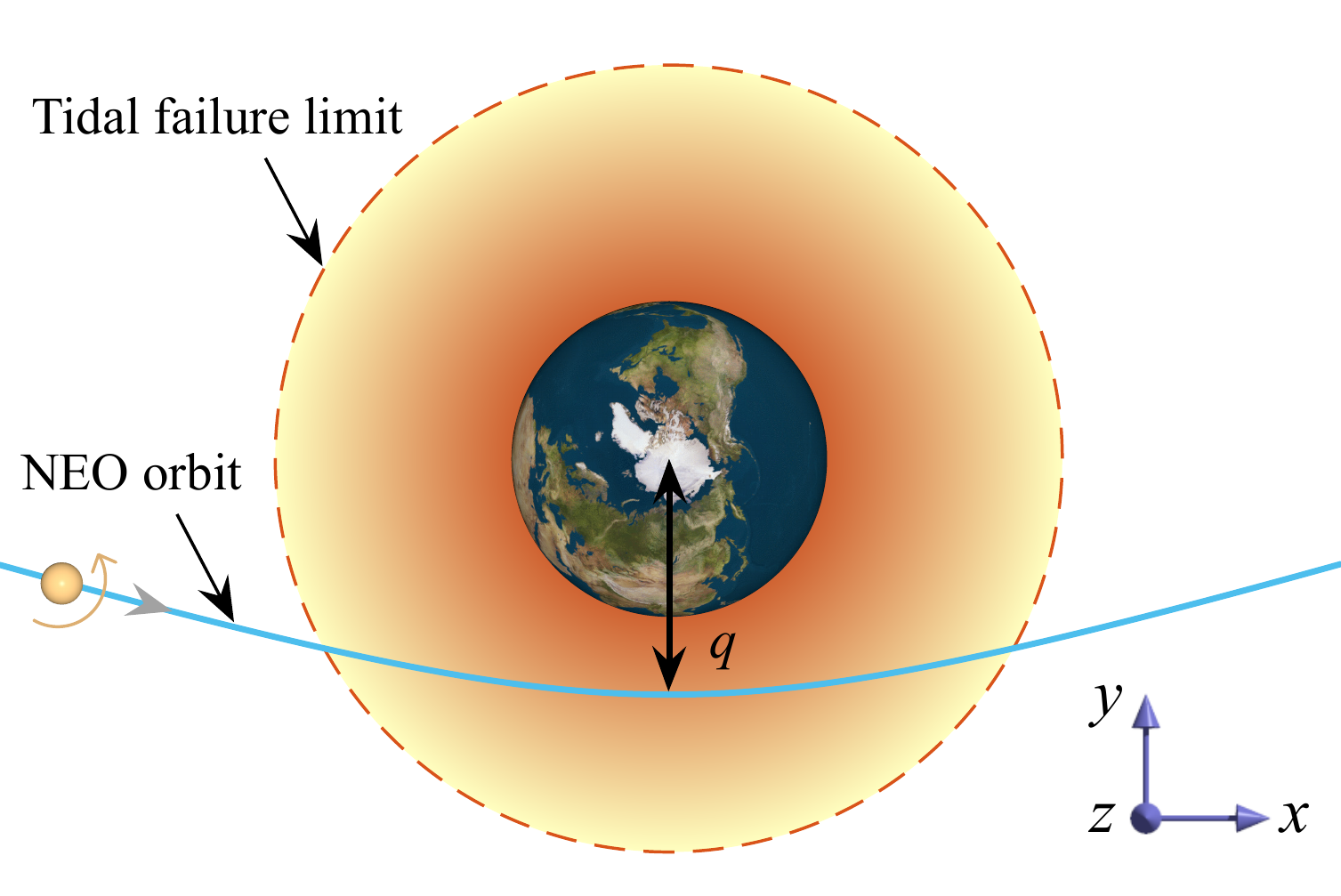}
    \caption{Representation of the kind of close Earth encounters considered in this study. The tidal failure region is given as the orange area surrounding the Earth.}
    \label{f:orbitdiagram}
   \end{figure}


  \begin{figure*}
    \centering
    \includegraphics[width = 16 cm]{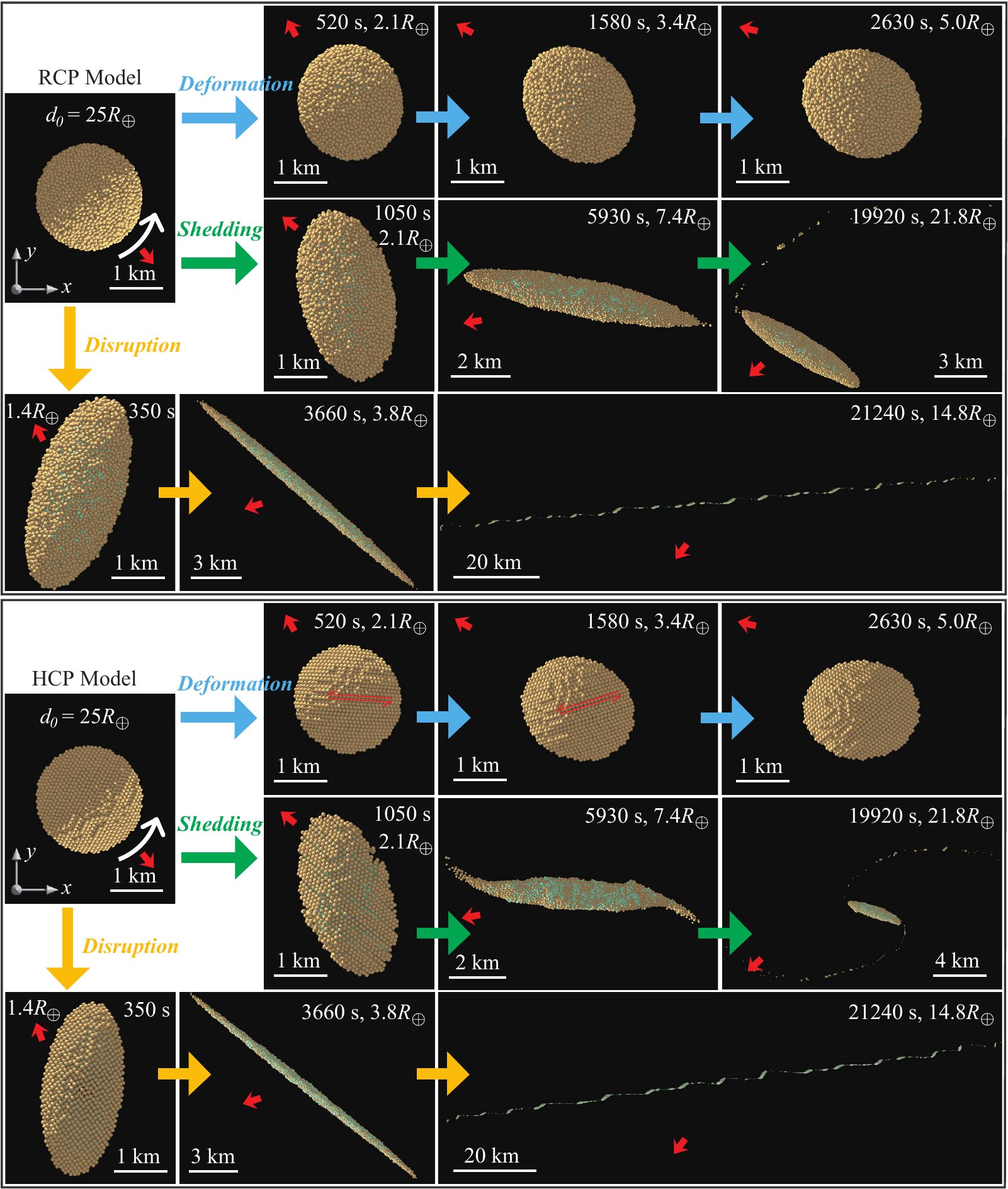}
    \caption{Snapshots of tidal encounter processes for the three outcome classes, ``deformation'' with $V_\infty=10$ km/s and $q=1.9R_\oplus$, ``shedding'' with $V_\infty=6$ km/s and $q=1.6R_\oplus$, and ``disruption'' with $V_\infty=2$ km/s and $q=1.3R_\oplus$, of the RCP (upper frame) and HCP (lower frame) models. The line of sight is in a direction perpendicular to the orbital plane (the same as Fig.~\ref{f:orbitdiagram}), and the light rays are ejected from the Earth to illustrate the Earth's direction (as indicated by the chunky red arrows in each snapshot). The top-left snapshot in each frame shows the start of each run, where the rubble pile rotates in a prograde direction. Time proceeds from left to right in each row. The values of time relative to the perigee and distance to the Earth's center are indicated on the top of each snapshot. Particles of the rubble pile at the surface are colored in yellow and interior ones are colored in green.  The thin red lines with arrows in the two middle snapshots of the HCP “deformation” case highlight the sliding motion between the two hemispheres of the HCP packing. \newline(An animation of each case shown in this figure is available.)}
    \label{f:snapshot}
  \end{figure*}
  
 We primarily consider close Earth encounters as examples to investigate tidal processes in this study.  We note that the tidal forces can be scaled with the mass and density of the primary body and the density of the body performing the flyby. Furthermore, the encounter velocity can be normalized to the escape velocity of the primary body. Therefore, our simulation results also provide a database for predicting tidal encounter outcomes for different conditions. 
 
 Figure \ref{f:orbitdiagram} illustrates the close Earth encounter scenario. The simulated rubble pile approaches the Earth on different hyperbolic orbits, which are defined by the encounter velocity at infinity, $V_\infty$, and the perigee distance, $q$. The theoretical tidal failure limiting distance to the Earth ($M_\oplus = 5.97\times 10^{24}$ kg, $R_\oplus = 6378$ km), $d_{\rm limit}$, for a cohesionless rubble pile with a friction angle of $18^\circ$, a prograde rotation of 4.3 h, and a bulk density of 2.47 g/cc, is about 2.5$R_\oplus$ \citep{Holsapple2006}. The distribution of $V_\infty$ for close hyperbolic encounters with the Earth ranges from 0 km/s to 40 km/s and peaks at $\sim$10 km/s \citep{Walsh2006}. Therefore, the perigee distance is set to range from 1.1$R_\oplus$ to 2.5$R_\oplus$, and the velocity at infinity is set to range from 0 km/s (parabolic orbit) to 20 km/s in our tests. The simulated rubble piles are taken to start at $d_0=10d_{\rm limit}$ from the Earth, which is large enough to ensure that Earth’s perturbations are negligible at the outset. Each run ends when all the tidal-induced fragments (or the reshaping rubble piles) have settled down to stable states, which typically occurs 1--2 days after perigee.

\section{Results}
\label{s:results}

   \begin{figure}
    \centering
    \includegraphics[width = 8.85 cm]{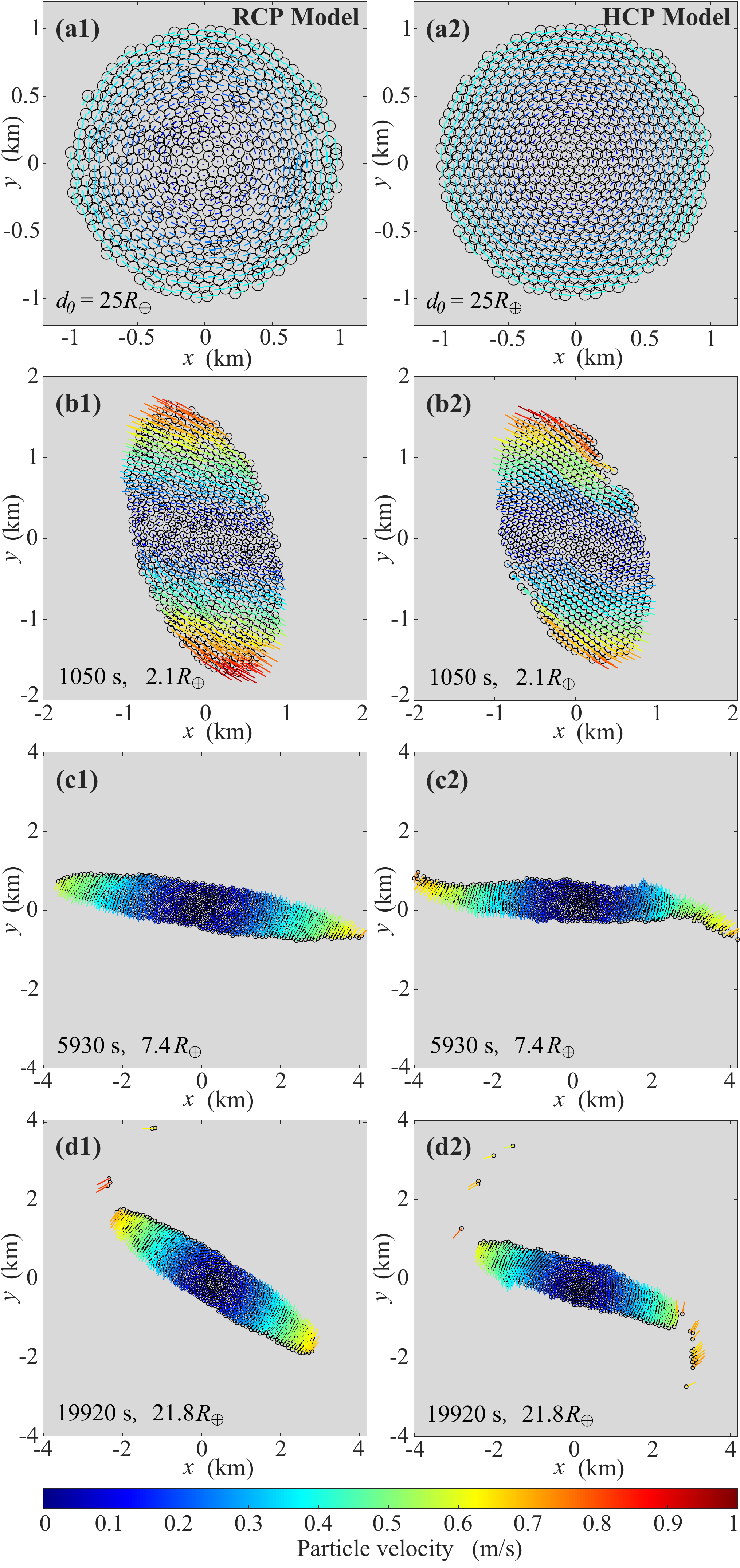}
    \caption{Particle velocity distribution over a cross-section parallel to the orbital plane (i.e., perpendicular to the maximum moment of inertia axis) at different stages of the encounter process for the RCP (left column) and HCP (right column) models with $V_{\infty}=6$ km/s and $q=1.6R_\oplus$ (i.e., the ``shedding'' case shown in Fig.~\ref{f:snapshot}). The particle velocity is computed with respect to the mass center of the rubble pile. The time relative to the perigee and distance to the Earth's center are given at the bottom of each plot. The colors and directions of the short lines correspond to the magnitudes and directions of the velocity vectors, respectively. }
    \label{f:velocityevolution}
   \end{figure}

 A tidal encounter can lead to surface motion and reshaping, mass shedding, and disruption of the progenitor. Figure \ref{f:snapshot} shows the snapshots of representative examples of our tidal encounter simulations, one using an RCP model, the other using an HCP model with the same material parameters, ($\mu_S$, $\beta$) = (0.2, 0.3), and the same particle resolution, $N\sim1$0,000. The three main tidal encounter outcome classes are presented for the two models. The ``deformation'' event (encounter conditions: $V_\infty=10$ km/s, $q=1.9R_\oplus$) leads to some changes in the shape and spin rate of the object without mass loss. The ``shedding'' event ($V_\infty=6$ km/s, $q=1.6R_\oplus$) leads to marginal mass loss at the extremities of the progenitor, in addition to shape and spin rate changes. Finally, the ``disruption'' event ($V_\infty=2$ km/s, $q=1.3R_\oplus$) leads to the full break-up of the progenitor into smaller separate pieces with various shapes and spin states.

 To account for all the effects caused by tidal encounters,  we analyze the outcomes of our simulations in terms of the largest remnant's mass, $M_\mathrm{lr}$,  elongation, $\varepsilon_\mathrm{lr}=1-(\alpha_1+\alpha_2)/2$ (where $\alpha_1$ and $\alpha_2$ are the short-to-long and medium-to-long axis ratios, respectively), and spin period, $P_\mathrm{lr}$, as a function of the periapse distance $q$ and the speed at infinity $V_\infty$. The largest remnant is simply the progenitor if no mass loss occurs. 

 The following subsections present the detailed tidal encounter outcomes for every case and highlight the factors that affect the tidal evolution of rubble-pile bodies.

\subsection{Tidal encounter processes and outcomes with SSDEM}
\label{s:results:outcomes}

 Because the RCP model is the most natural representation of a rubble-pile interior, we start by investigating this model using the SSDEM and present the outcomes of tidal encounters in detail. 

 The left column of Fig.~\ref{f:velocityevolution} shows the particle velocity evolution of the RCP model with $\phi=18^\circ$ in a ``shedding'' event (other types of events show similar behaviors). In the beginning, the velocity distributes following a linear function of the distance from the center as the rubble pile is constantly rotating (see Fig.~\ref{f:velocityevolution}a1). When passing by the Earth, parts of the RCP model that are closer to the Earth are attracted more strongly by gravity of the Earth than the other parts that are farther away. The disparity is larger than the force of effective gravity (combined with the rotational effects) holding the rubble pile together when it enters the tidal failure limit. The near and far parts of the rubble pile are pulled apart from each other, leading to increases in radially outward velocities (see Figs.~\ref{f:velocityevolution}b1). After leaving the tidal failure region, the RCP rubble pile is continuously stretched until the outward velocities are dampened by gravitational interactions, collisions, and frictions between particles (see Fig.~\ref{f:velocityevolution}c1, d1). These factors also help the remnants to establish new stable structures. 

   \begin{figure}
    \centering
    \includegraphics[width = 8.9 cm]{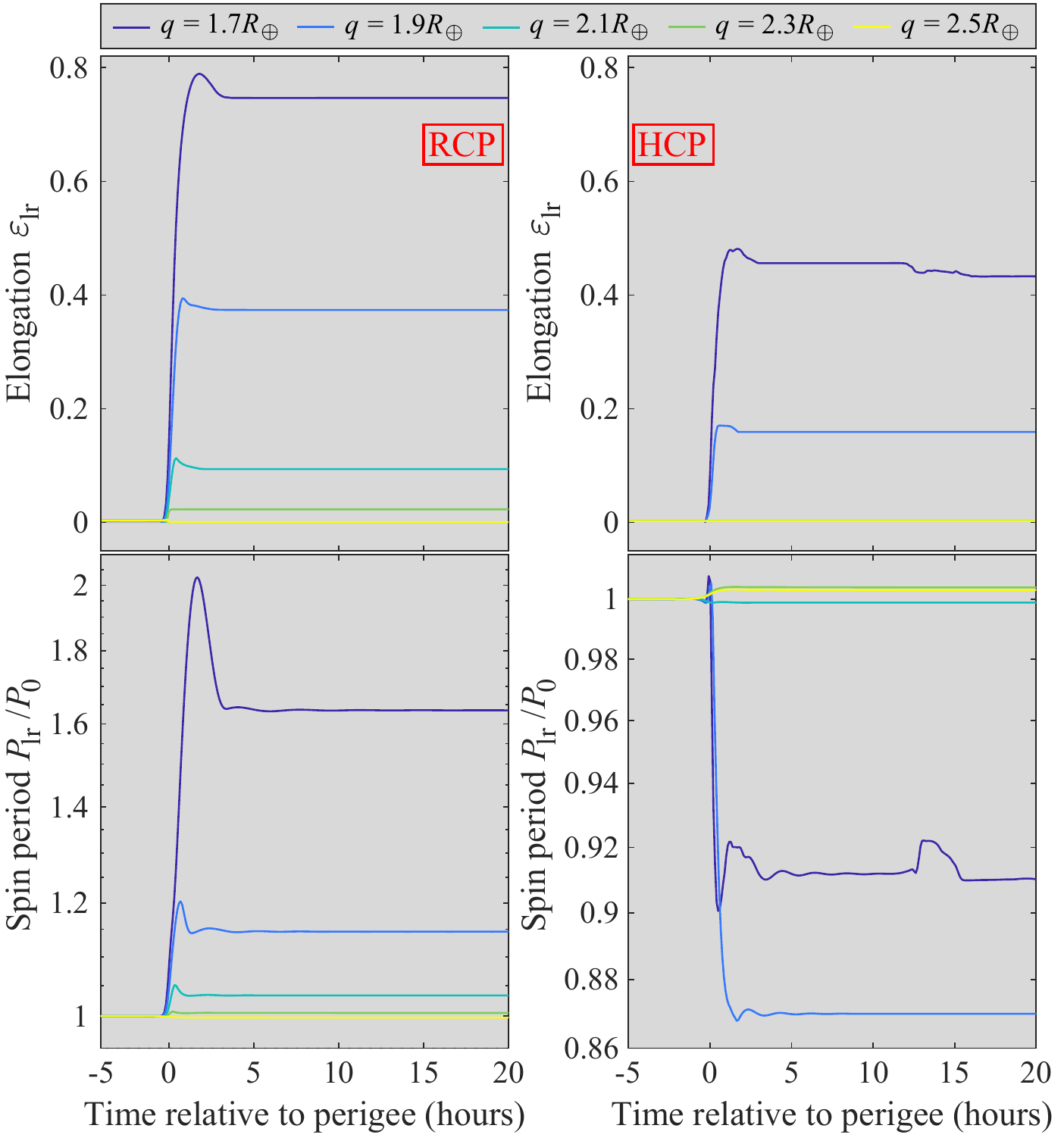}
    \caption{Elongation and spin period evolution of the largest remnants during tidal encounters for the RCP (left column) and HCP (right column) models with $V_{\infty}=5$ km/s and different $q$. The colors of the curves represent the results with different $q$ as indicated in the top legend. }
    \label{f:tidalprocess_vinf5}
   \end{figure}

  \begin{figure}
    \centering
    \includegraphics[width = 8.9 cm]{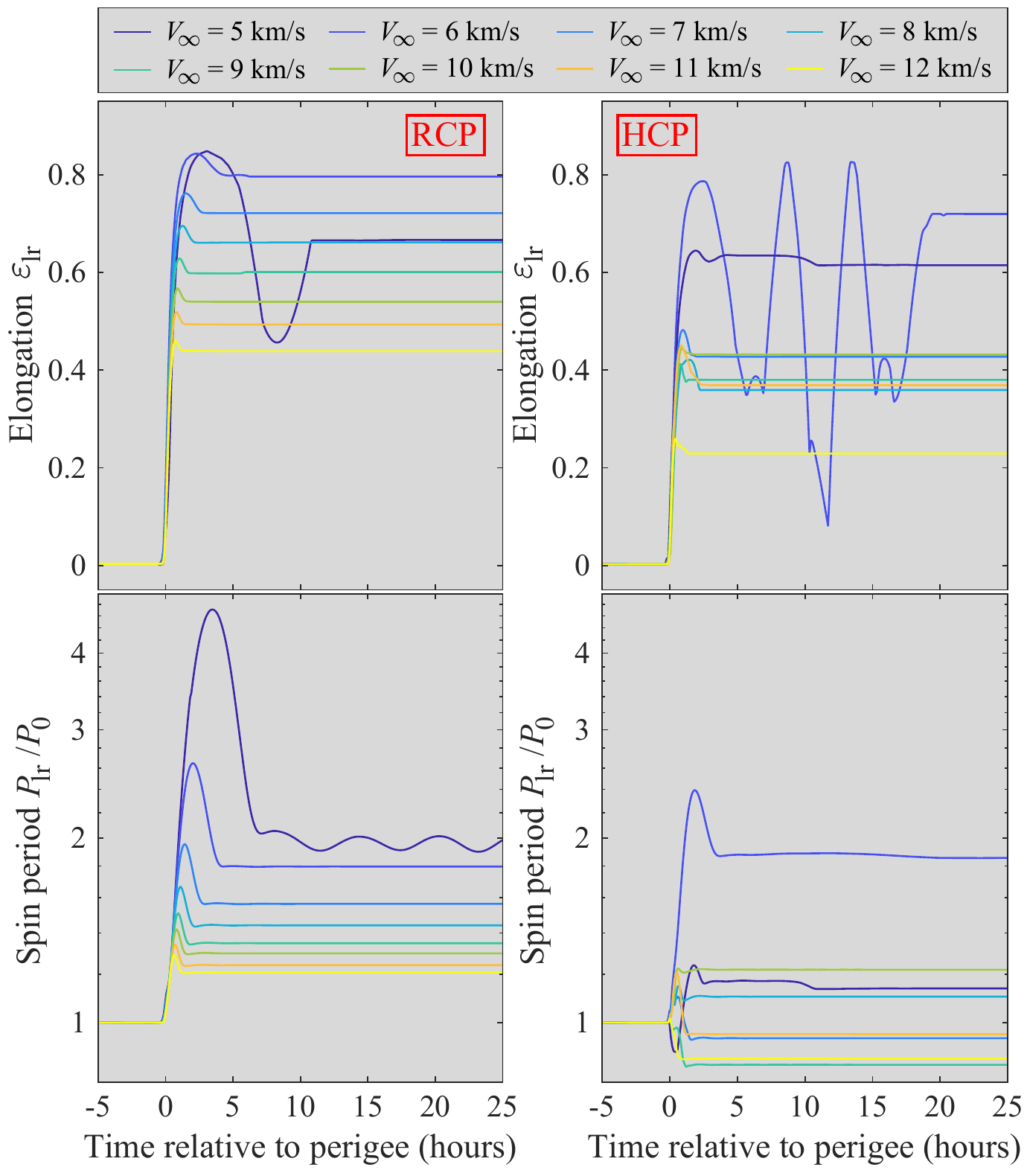}
    \caption{Similar to Fig.~\ref{f:tidalprocess_vinf5} but for $q=1.6R_\oplus$ and different $V_{\infty}$.}
    \label{f:tidalprocess_q16}
  \end{figure}

  \begin{figure*}
   \centering
   \includegraphics[width = 16.2 cm]{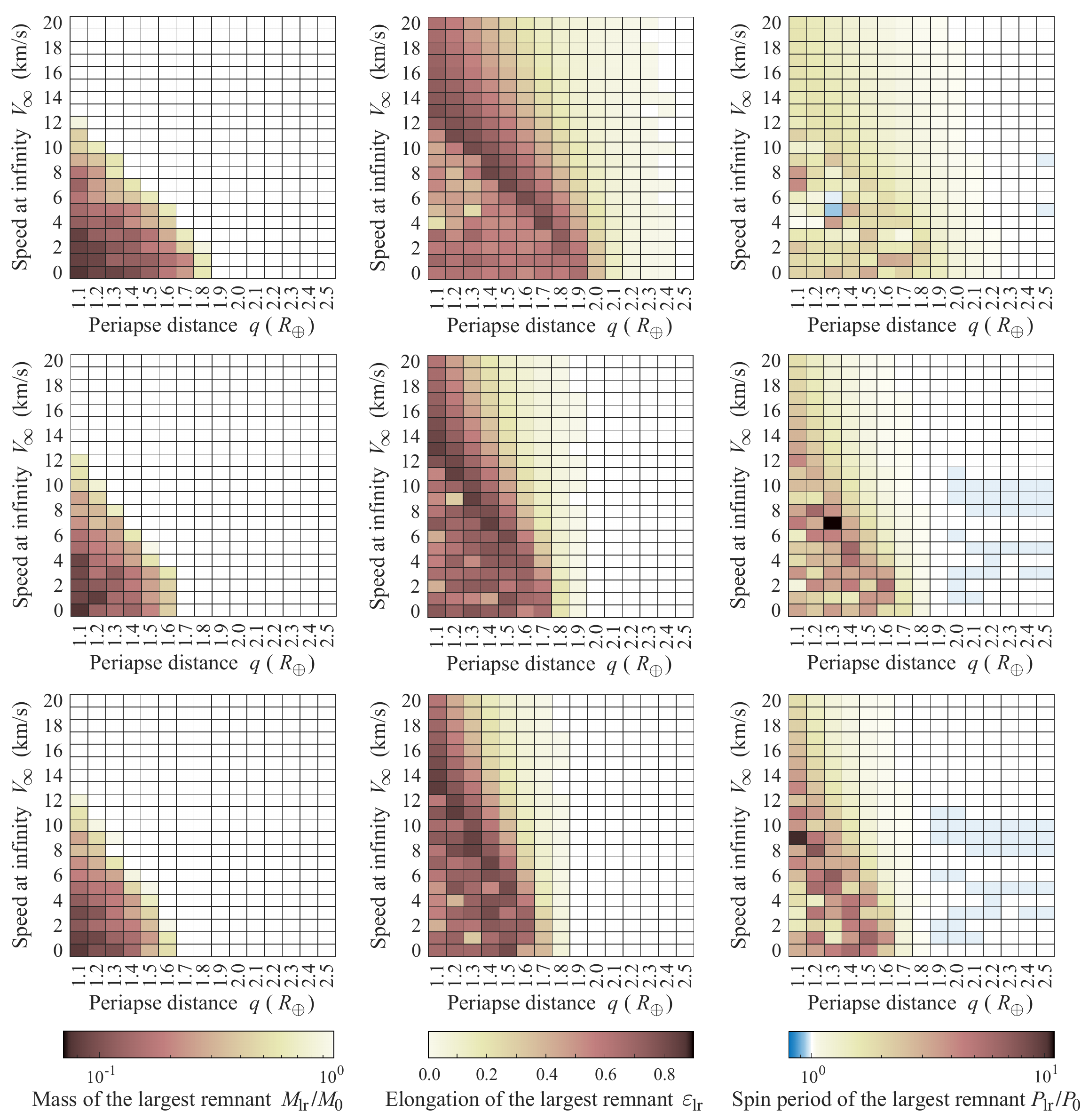}
   \caption{Tidal fragmentation outcomes for the RCP model with different friction angles, where $\phi=18^\circ$ for the top row, $\phi=27^\circ$ for the middle row, and $\phi=32^\circ$ for the bottom row. The three columns present the mass ratio, $M_\mathrm{lr}/M_0$, the elongation, $\varepsilon_\mathrm{lr}$, and the spin period normalized by the initial spin period, $P_\mathrm{lr}/P_0$, of the largest remnant as a function of the speed at infinity, $V_\infty$, and the periapse distance, $q$, of the Earth-encountered progenitors, respectively. }
   \label{f:frictioneffect}
   \end{figure*}

 The left columns of Figs.~\ref{f:tidalprocess_vinf5} and \ref{f:tidalprocess_q16} present the shape and spin period evolution of the largest remnants of the RCP model with $\phi=18^\circ$ in some ``deformation'' and ``shedding'' events (the mass loss can be inferred from the top row of Fig.~\ref{f:frictioneffect}). The values of the elongation and spin period were calculated using a clump of particles that eventually form the largest remnant for a given encounter (hereafter, ``bound particles''). For ``shedding'' events, such as the case of $V_\infty=5$ km/s shown in Fig.~\ref{f:tidalprocess_q16}, in which the largest remnant is a fraction of the progenitor and most of particles at the extremities of the progenitor escape, the elongation increasing rate in the beginning of structural failure is smaller than that of the other cases. The bound particles' orbiting movements and collisions involved in reaccumulation processes also bring some vibrations in the elongation evolution. As the spin state is dramatically modified, the largest remnant can become a tumbling rotator (i.e., non-principal axis rotation; e.g., in Fig.~\ref{f:tidalprocess_q16}, the spin period of the case with $V_\infty=5$ km/s shows a mode of short periodic oscillations). For ``deformation'' events, the bound particles are all particles composing the progenitor. The body is first elongated as a result of tidal forces and then, when the tidal effects become negligible, it shrinks until the contacts and frictions between the constituting particles are capable of maintaining a stable configuration. There is a clear trend that the rubble pile becomes more elongated and spins slower for a closer or slower encounter, as the tidal forces are stronger or have longer time to exert their influence. 

 Figure \ref{f:frictioneffect} presents the mass, elongation, and spin period of the largest remnant as a function of the close approach distance $q$ and the encounter speed at infinity $V_\infty$ for RCP progenitors with different friction angles. Due to the intrinsic material shear strength and the short flyby time, the rubble-pile object can only be tidally disrupted at a distance notably lower than the theoretical tidal failure limiting distance (e.g., the theoretical limiting distance for $\phi=18^\circ$ is $\sim2.5R_\oplus$, while our simulation results show mass loss occurs when $q < 1.9R_\oplus$ for the lowest friction case). Since the tidal forces dramatically increase with a closer orbit, there is an obvious trend that the smaller the periapse distance $q$, the more efficient the encounter to modify the progenitor. For weak encounters, where the perigee distance is close to the limiting distance, only ``deformation'' events can be triggered and the rubble pile is slightly distorted to a prolate shape when passing by the Earth. The distortion of the object becomes more severe with a smaller perigee distance. When the perigee distance is close to $1.4R_\oplus$, the rubble pile is heavily distorted and disrupted by Earth’s tides. Similarly, with increases in the encounter time within the limiting distance, the progenitor is subject to more brutal disruption and is split into larger amounts of fragments for a smaller speed at infinity $V_\infty$. 

 As shown in the left column of Fig.~\ref{f:frictioneffect}, the mass of the largest remnant tends to monotonously decrease with a smaller $q$ or $V_\infty$ as the components can obtain larger tidal accelerations and escape from each other. In intensive ``disruption'' events, where $M_\mathrm{lr} \lesssim 0.1M_0$, tens of fragments with size similar to the largest remnant are produced concurrently (the distribution is analogous to the SL9 fragment trains; see the ``disruption'' event shown in Fig.~\ref{f:snapshot} for an example).

 The elongation and spin period of the largest remnant show more complicated dependencies on the encounter conditions. From the middle column of Fig.~\ref{f:frictioneffect}, we observe that there is an optimal encounter configuration, in terms of both $q$ and $V_\infty$, that produces a largest remnant with the maximal elongation ($\varepsilon_\mathrm{lr}>0.8$). Considering both the left and middle columns, the maximum elongations occur under such circumstances where the rubble pile is about to lose mass. In these cases, the largest remnant marginally maintains its structural stability (see Fig.~\ref{f:spin_path} and discussions in Sect.~\ref{s:results:continuum} below). A more severe tidal encounter can break this extremely elongated shape apart, while a more mild tidal encounter is inadequate to reshape the object into such an extreme shape.

 The spin period evolution of the largest remnant is associated with its elongation evolution. As shown in the left columns of Figs.~\ref{f:tidalprocess_vinf5} and \ref{f:tidalprocess_q16}, the increases in the spin period lag behind the increases in the elongation. The rapidly rotational deceleration indicates that the angular momentum increment due to tidal torques is insufficient to maintain the initial rotation speed for the RCP model. Therefore, the RCP model is usually spun down as a result of being more elongated by conservation of angular momentum. However, in the top-right panel of Fig.~\ref{f:frictioneffect}, one can observe a spin-up of the largest remnant ($P_\mathrm{lr}<P_0$) at $q=1.3R_\oplus$ and $V_\infty=5$ km/s. This is due to the fact that in such encounter conditions, the rubble pile is disrupted (see the top-left panel of Fig.~\ref{f:frictioneffect}) and in this particular case, the disruption leads to a near-spherical largest remnant ($\varepsilon_\mathrm{lr}\sim0.2$; see the top-middle panel of Fig.~\ref{f:frictioneffect}) and results in a decrease of its spin period. In contrast, for most of the ``disruption'' cases, the largest remnant is highly elongated and its spin period is much larger than the initial state.

\subsection{Effect of material shear strength}
\label{s:results:material_effect}

 The material shear strength in our SSDEM is mainly controlled by the tangential friction coefficient, $\mu_S$, and the rotational resistance shape parameter, $\beta$. With the resistance forces/torques brought by $\mu_S$ and $\beta$, spherical particles in our SSDEM can capture part of the behavior of realistic particles with irregular shapes in the way those particles resist relative sliding and rotational movement \citep{Zhang2017, Zhang2018}. We test the effect of material shear strength by conducting tidal encounter simulations using the RCP model with three friction angles, namely, $\phi=18^\circ$, $27^\circ$, and $32^\circ$, corresponding to $(\mu_S,~\beta)=(0.2,~0.3)$, $(0.6,~0.5)$, $(1.0,~0.8)$, respectively.
 
  \begin{figure}
   \centering
   \includegraphics[width = 8.9 cm]{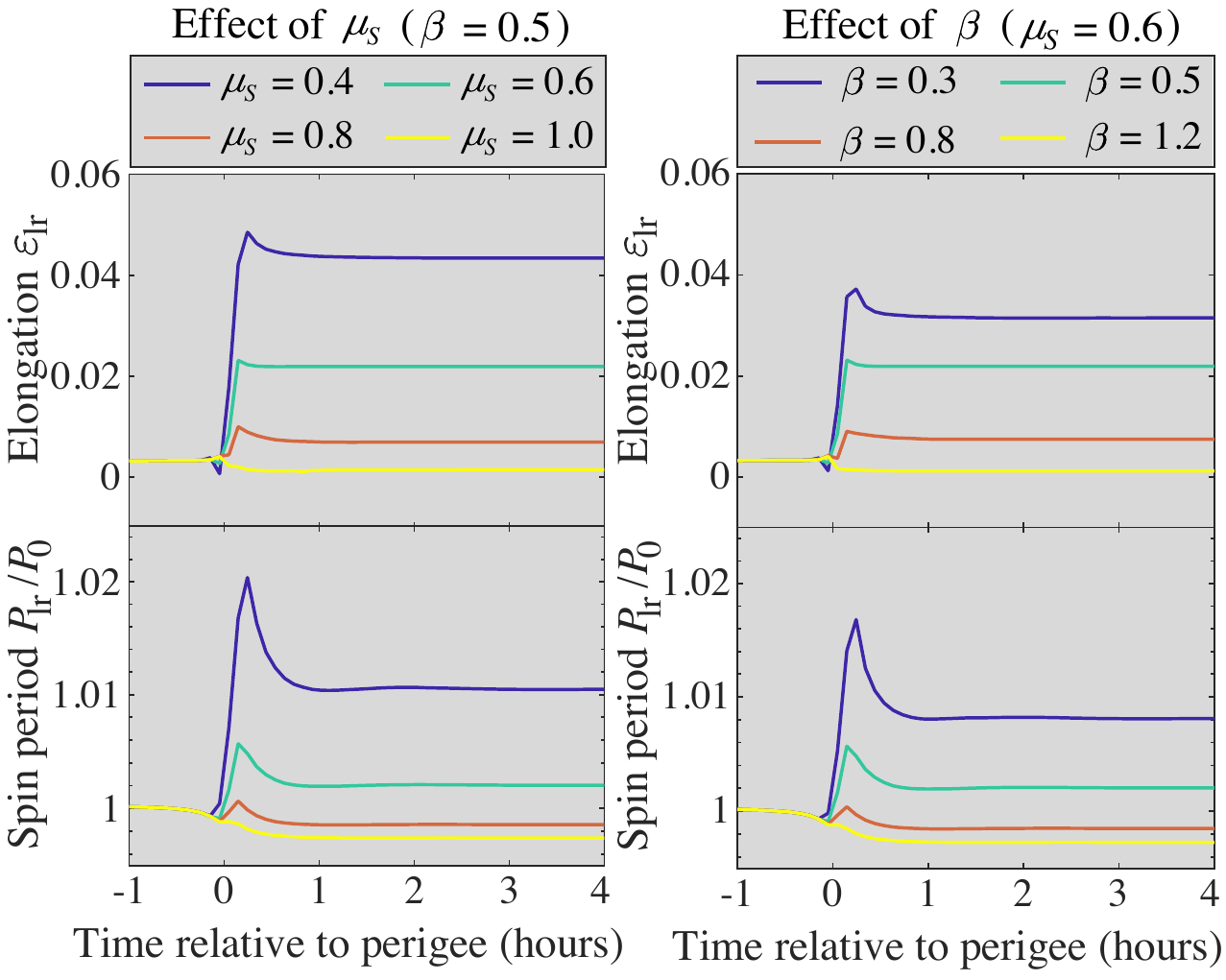}
   \caption{Elongation and spin period evolution of the largest remnants during tidal encounters for the RCP model with $V_\infty = 10$ km/s and $q=1.9R_\oplus$ (mild deformation) for different tangential friction coefficients $\mu_S$ (left column) and shape parameters $\beta$ (right column).}
   \label{f:rollingfrictionv10}
  \end{figure}  
   
  \begin{figure}
   \centering
   \includegraphics[width = 8.9 cm]{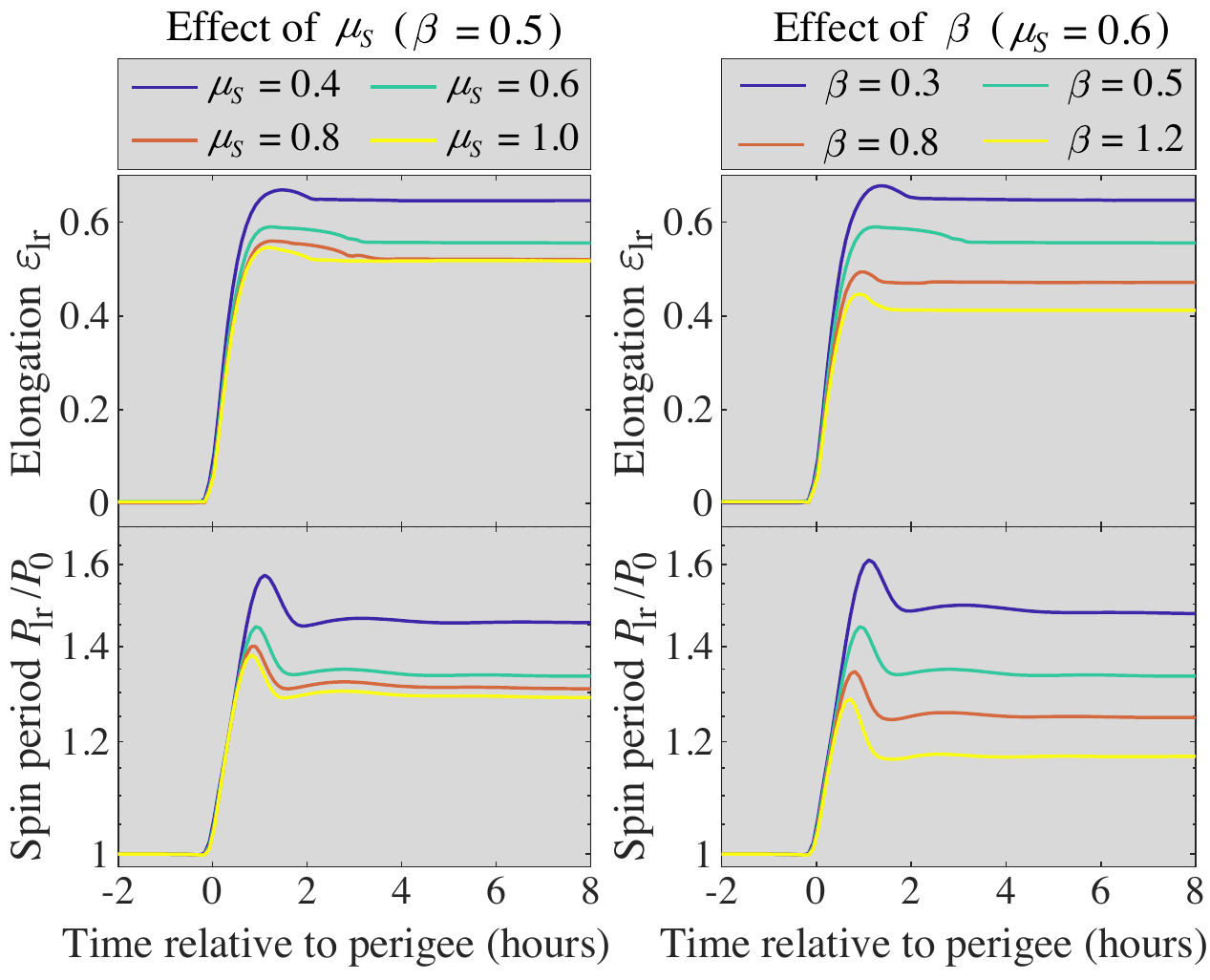}
   \caption{Similar to Fig.~\ref{f:rollingfrictionv10} but for $V_\infty = 6$ km/s and $q=1.6R_\oplus$ (marginal shedding).}
   \label{f:rollingfrictionv6}
  \end{figure}  

  \begin{figure}
   \centering
   \includegraphics[width = 8.9 cm]{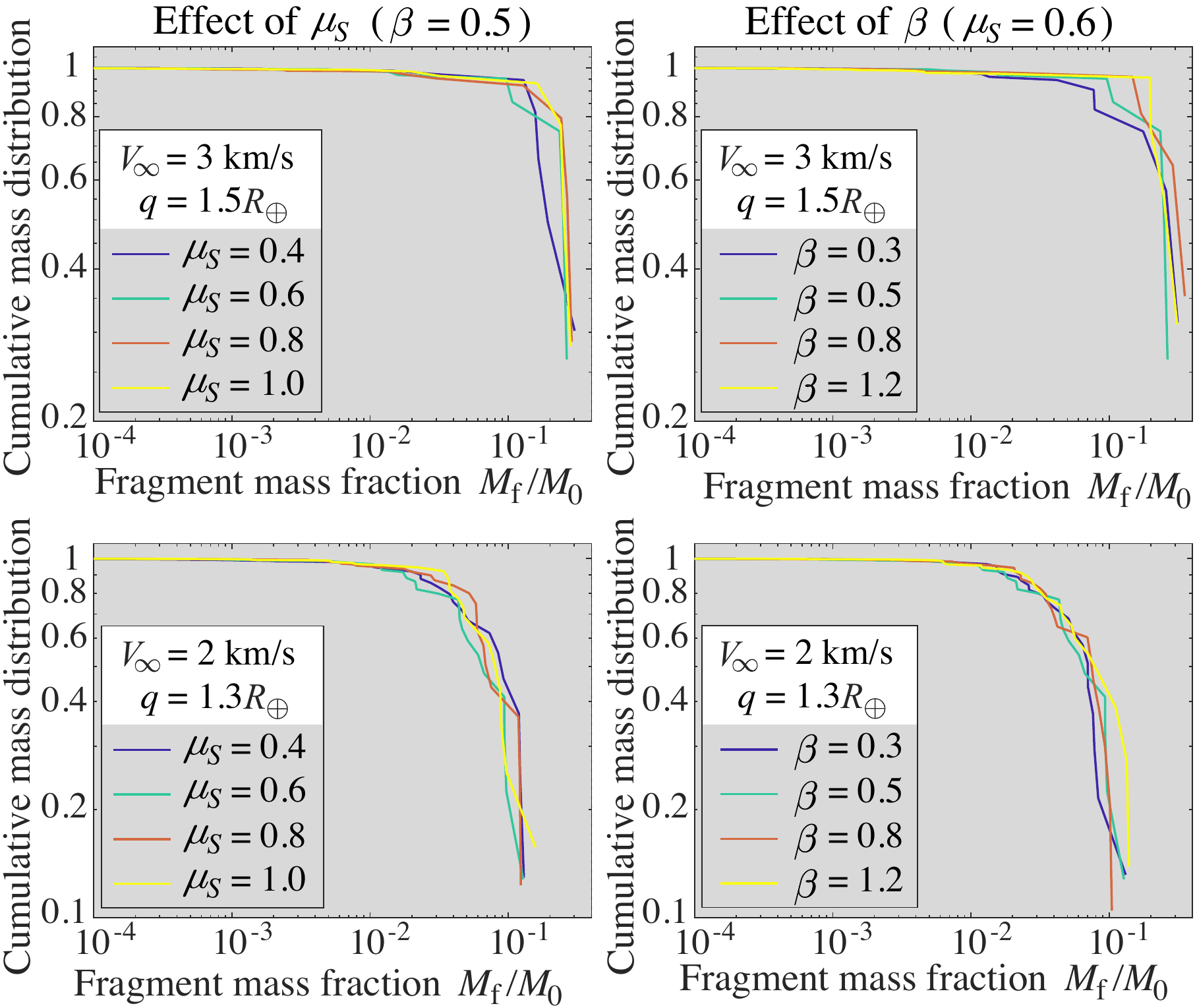}
   \caption{Cumulative fragment mass distributions for the RCP model with $V_\infty = 3$ km/s and $q=1.5R_\oplus$ (top; moderate disruption), and $V_\infty = 2$ km/s and $q=1.3R_\oplus$ (bottom; catastrophic disruption), at the end of the SSDEM simulations ($\sim$85$R_\oplus$ away from the Earth). The colors denote the results with different tangential friction coefficients $\mu_S$ (left column) and shape parameters $\beta$ (right column).}
   \label{f:rollingfrictionv2v3}
  \end{figure}  
   
  \begin{figure*}
   \centering
   \includegraphics[width = 16.2 cm]{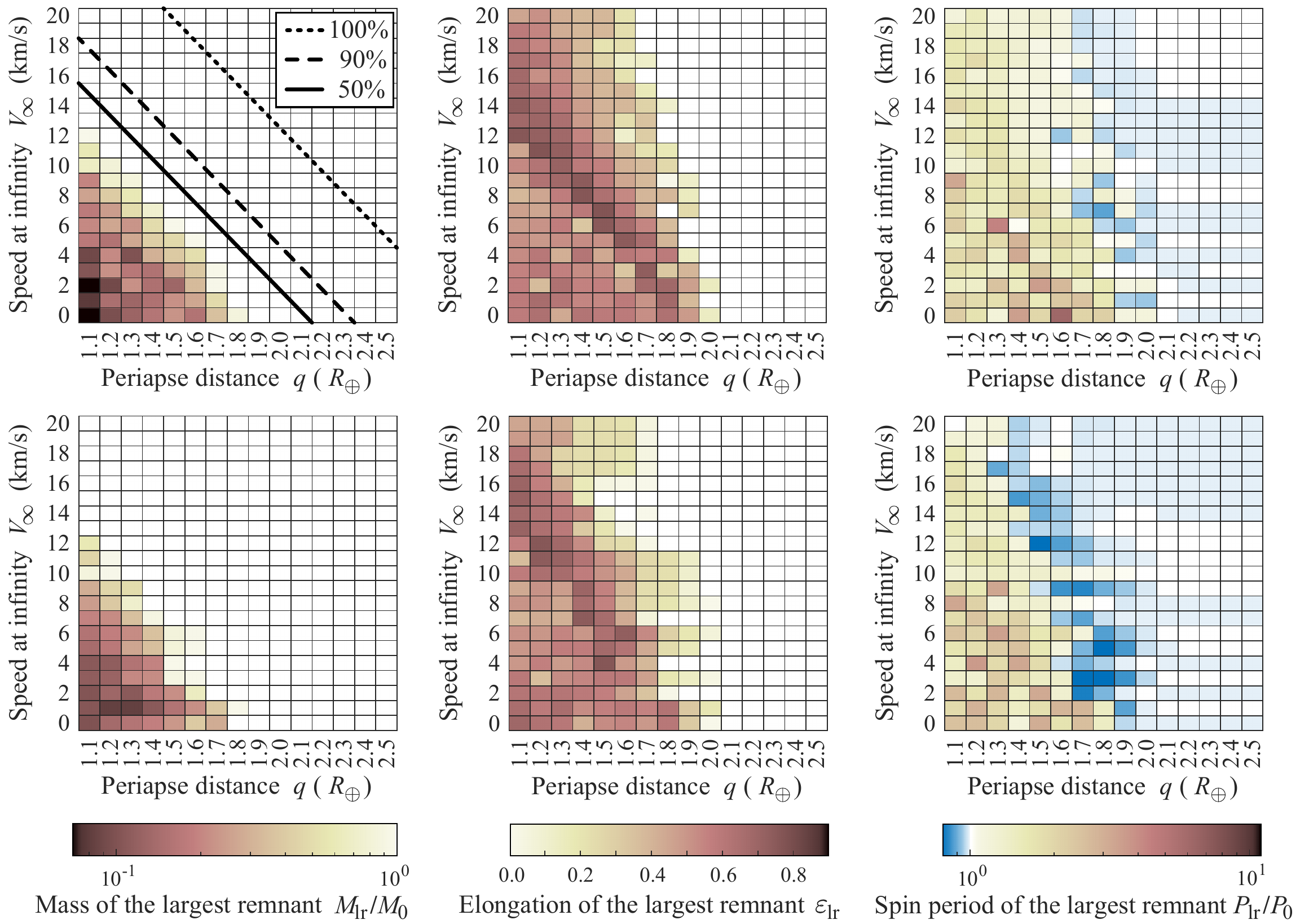}
   \caption{Similar to Fig.~\ref{f:frictioneffect} but for the HCP model with different resolutions, where $N=2953$ for the first row and $N=11577$ for the second row. The three lines plotted on the top-left panel indicate the tidal disruption mass loss outcomes of previous simulations using the HSDEM \citep[based on Fig.~1 in ][]{Schunova2014}, where above the ``100\%'' line the progenitor has no mass loss, below the ``90\%'' line more than 10\% of the progenitor's original mass is lost, below the ``50\%'' line the mass loss is larger than half of the original mass $M_0$ (see Sect.~\ref{s:ComparingHS}).}
   \label{f:resolutioneffect}
  \end{figure*}  
   
 As shown in Fig.~\ref{f:frictioneffect}, with lower shear strength (i.e., angle of friction), the RCP model begins to lose mass, reshape and experience spin changes at larger periapse distances. This trend appears continuous from the lowest to the highest shear strength cases. Based on the static continuum theory of \cite{Holsapple2006}, the theoretical tidal failure limiting distance for the simulated rubble piles with the three considered friction angles is $2.5R_\oplus$, $2.1R_\oplus$, and $1.9R_\oplus$, respectively. Above the corresponding limit, the rubble piles are impervious to the tidal effects and can preserve their original shapes and rotation states. The results shown in Fig.~\ref{f:frictioneffect} are consistent with the theoretical analyses. Although some light blue patches, presented on the right side of the spin period results in Fig.~\ref{f:frictioneffect}, indicate that the rubble piles are slightly spun up by the tidal torques at the distance above the tidal failure limit, the increments are too small (on the order of $10^{-4}P_0$) to reveal systematical pattern and are very sensitive to tiny changes in the shape. 

 However, according to the left column of Fig.~\ref{f:frictioneffect}, as the periapse distance $q$ decreases, the influence of the shear strength on the mass loss situation appears to become less pronounced. This can be explained by considering that the rubble pile is heavily stretched by the tidal forces and particles start to lose contacts with others in these extremely close encounters, and therefore the interparticle friction becomes ineffective. Through analyzing the force chain evolution of rubble-pile bodies in these extremely close encounter simulations, we confirmed that the coordination number (i.e., the average number of contacts for each particle) drops to almost zero near the perigee and resumes to the normal level of a compacted packing (e.g., $\sim$4 for the RCP model) during the post-encounter reaccumulation stage. For a smaller $q$, the zero-contact stage lasts longer, and the role of contact friction is less important during the course of the breakup.

 This contact-loss behavior also affects the effect of shear strength in different kinds of encounter scenarios. We run extra simulations with four encounter scenarios to reveal the independent influence of the material parameters $\mu_S$ and $\beta$. Figures \ref{f:rollingfrictionv10}, \ref{f:rollingfrictionv6}, and \ref{f:rollingfrictionv2v3} show tidal encounter outcomes for RCP models corresponding to mild deformation, marginal shedding, moderate and catastrophic disruption, respectively. For each condition, either different $\mu_S$ and a fixed $\beta$, or different $\beta$ and a fixed $\mu_S$, are used. Comparing the curves for different $\mu_S$ and for different $\beta$ in Figs.~\ref{f:rollingfrictionv10} and \ref{f:rollingfrictionv6}, we can observe that the effect of the sliding resistance $\mu_S$ is more pronounced in mild deformation events, while the effect of the rotational resistance $\beta$ becomes dominant in marginal shedding events. In moderate disruption events (see the top row of Fig.~\ref{f:rollingfrictionv2v3}), the fragment mass distribution does not show clear and systematic trends for different $\mu_S$, while the mass fraction of the massive fragments increases with $\beta$. This implies that, for more intensive tidal encounters (from mild deformation to moderate disruption), enhancing the rotational resistance between particles has more influence on the tidal encounter outcome than increasing the sliding resistance. In catastrophic disruption events, which involve many separation, collision and reaccumulation events between various components of the rubble pile, the fragment distribution is the outcome of chaotic processes and the material friction parameters play little roles (see the bottom row of Fig.~\ref{f:rollingfrictionv2v3}).

 Nevertheless, since the material strength is important for maintaining the structural stability of a rubble pile, the shape and spin of the largest remnant generally closely depend on the friction angle in all encounter scenarios, as shown in Fig.~\ref{f:frictioneffect}. For ``deformation'' events with the same encounter conditions, as the friction hinders the relative movement between particles, the largest remnant is less elongated and spins faster with higher shear strength.  For ``shedding'' and ``disruption'' events, the largest remnant generally has a more elongated shape and rotates slower with higher shear strength. This is because a rubble pile with lower shear strength needs to rotate fast enough so that it does not shrink. This effect is emphasized for elongated remnants (see Fig.~\ref{f:lr_distribution} and Sect.~\ref{s:results:continuum} below for details).

 \subsection{Effect of internal packing}
\label{s:results:packing}
   
 The actual internal packing of a rubble pile remains a big question since no direct measurement of the internal structure of an asteroid has ever been performed. Moreover, it is not clear whether there is a more typical internal configuration of a rubble pile formed through reaccumulation after a parent body's disruption or whether this process can lead to a wide diversity of internal structures \citep{Michel2015, Michel2020}. However, an RCP model may be a better representation than an HCP model, as it is more natural to reach than an HCP when particles are assembled together and shacked. As the geometric interlocking is largely affected by the arrangement of spherical particles and the internal packing plays a significant role in the structural stability and dynamical evolution of rubble-pile bodies \citep{Zhang2017}, while past studies of tidal encounters only considered the HCP model \citep[e.g.,][]{Schunova2014,DeMartini2019}, comparing the outcomes of tidal encounters using both the HCP and RCP models is important to understand the effect of internal structure of rubble piles. 
 
 Figure \ref{f:frictioneffect} (top row) and Fig.~\ref{f:resolutioneffect} (bottom row) show the results for the RCP and HCP models, respectively, using the same resolution ($N\sim1$0,000) and material parameters ($\mu_S=0.2$, $\beta=0.3$). The HCP model starts breaking-up at smaller periapse distances than the RCP one. The RCP model is more severely subject to tidally-induced reshaping at larger $q$ and larger $V_\infty$ than the HCP model. Generally, while a continuous trend of the state of the largest remnant in the ($q$, $V_\infty$) parameter space in the non-disruptive cases can be easily identified for the RCP model, there is no such obvious trend for the HCP one and in some cases, the HCP model can be spun up by the tidal forces without any strong reshaping. 
 
 These differences can be explained by analyzing the detailed tidal encounter processes of these two models. As shown in Fig.~\ref{f:snapshot}, for each tidal encounter outcome class (i.e., ``deformation'', ``shedding'', and ``disruption''), the evolution under the action of tidal forces and final states for the HCP model are very different than those for the RCP model. This is because the effects of the tidal forces acting on an HCP model are constrained by the specific interlocking geometry imposed by this kind of packing (i.e., the so-called ``cannonball stacking''). Unlike the homogeneous deformation of the RCP model, the HCP model exhibits some visible cracks in the beginning of tidal failure. Given that the motion of internal particles in the HCP structure is restricted by their neighbors, the formation of cracks is necessary to distort this structure. There are only two modes allowing for the deformation of the HCP structure, namely, internal particle detachment mode or surface particle movement mode. The former mode is observed for the ``deformation'' case shown in Fig.~\ref{f:snapshot}. As indicated by the thin red arrows in the two middle snapshots, the tidal tensile stress pulls the HCP apart and the tidal torque leads the resulting two parts to slide along each other, producing a prolate rubble pile (see Supplementary Video 4). The HCP configuration can be preserved in this case. For more severe encounters, e.g., the ``shedding'' case shown in Fig.~\ref{f:snapshot} and Supplementary Video 5, both modes are active and the particle distribution is subject to heavily rearrangement. Compared with the RCP model whose internal particles can move more freely and respond to the tidal force with a radially symmetrical velocity distribution (see Fig.~\ref{f:velocityevolution}b1), the internal particles in the HCP model mainly move along their corresponding particle band in a direction close to the tidal force direction (see Fig.~\ref{f:velocityevolution}b2). The gaps among these particle bands are increased by the tidal force, so the internal particles can move outward to break this close packing. In ``disruption'' events, for example, the case shown in Fig.~\ref{f:snapshot} and Supplementary Video 6, the hexagonal configuration is completely destroyed and the fragmentation outcomes are similar to the outcomes of the RCP model.
 
 As the breakage of the HCP structure takes some time, the deformation of the HCP model lags behind the RCP model and the interlocking of the internal particles slows down the shape change rate of the HCP model under the same encounter conditions (see Figs.~\ref{f:tidalprocess_vinf5} and \ref{f:tidalprocess_q16}). These effects are more effective in non-disruptive events as such weak encounters need more time to break the interlocking, which allows the HCP model to spin faster than the RCP one. In some ``deformation'' events, where the angular momentum increment imposed by the tidal torque exceeds the angular momentum needed for maintaining a constant rotation during deformation, the HCP model produces a fast rotating largest remnant with $P_\mathrm{lr}<P_0$. In ``shedding'' events, this fast spin also leads to more efficient mass shedding on the surface and at the tips of the HCP progenitor, compared to the RCP model (see the ``shedding'' cases in Fig.~\ref{f:snapshot}). For the RCP model, the spin-down rate caused by shape change surpasses the spin-up rate provided by the tidal torque, and therefore the tidal encounter only leads to the spin-down of the RCP progenitor in non-disruptive events (e.g., see the bottom-left panel in Fig.~\ref{f:tidalprocess_vinf5}).
 
 Due to the highly asymmetric particle arrangement in the HCP model, tidal effect can be reinforced when the direction of the particle bands is aligned with the tidal force direction. On the other hand, the HCP model can be stronger against tidal forces when these forces act in the same direction as the one for which the geometry of the particle arrangement prevents the particles to move. Therefore, the outcome of a tidal encounter for an HCP model depends sensitively on the precise internal configuration and orientation of the model towards the tidal force direction. This can explain why there is no continuous trend in the tidal encounter outcomes as a function of $q$ and $V_\infty$, as we find for the RCP model (see Figs.~\ref{f:tidalprocess_vinf5} and \ref{f:tidalprocess_q16}). 
 
 The effect of body orientation during tidal encounters has been reported in previous numerical studies. However, as these studies only used HCP models in their simulations, the influences of shape orientation and particle arrangement orientation were mixed together. For example, the study of \cite{DeMartini2019}, where the same SSDEM code {\it pkdgrav} and an HCP configuration based on a radar shape model were used to simulate asteroid Apophis's close encounter with the Earth in 2029, shows that different encounter orientation can result in very different changes in the asteroid's spin period, from large spin up to spin down, or no effect at all. Combined with our above analyses, the sensitivity of encounter outcomes to the body orientation may be mainly caused by the asymmetric characteristic of the HCP model, rather than the shape irregularity. For the homogeneous RCP model, the particle arrangement orientation has little effect and the body is unlikely to be significantly spun up by the encounter. To eliminate the uncertainties due to the special HCP configuration, a more precise estimation on the encounter outcomes of Apophis and the effect of body orientation can be obtained by using a random packing, which is left for future studies. 
 
\subsection{Effect of rubble-pile model resolution}
\label{s:results:resolution}

 Images returned by small-body missions reveal that these small bodies are covered by over billions of grains and boulders \citep[e.g.,][]{Walsh2019,Sugita2019}. Current computational power cannot afford simulations with such particle number levels. We have to reduce the particle resolution to model small bodies for practical reasons. Carrying out a study on the effect of resolution could help us to justify our simulation results and interpret the implications for real small bodies. Previous numerical studies showed that the tidal disruption outcomes do not strongly depend on the particle number in a range from $10^3$ to $10^5$ for random packing rubble-pile progenitors \citep{Movshovitz2012, Zhang2020}. \cite{Richardson2005} used the HSDEM and HCP models to test the structural stability of Jupiter's satellite Amalthea using a number of particles ranging from $N=2$ to $N=10^4$, and found that coarse configurations consisting of a small number of particles are more resistant to tidal disruption than fine configurations with more particles. Here, using the SSDEM, we consider two HCP models with different resolutions ($N=2953$ and $N=11577$) to test the resolution effect.

 Figure \ref{f:resolutioneffect} compares the results obtained with these two models. Generally, increasing by a factor of $\sim$4 the number of particles makes the HCP model harder to be disrupted and less deformable during tidal encounters, and spin faster in some cases. The discontinuity of the distribution of the largest remnant's state variables in the ($q$, $V_\infty$) parameter space is more pronounced in the high-resolution model. That is, all the characteristics of HCP configurations in response to the tidal encounter are magnified in the high-resolution model as the movement of its internal particles is more restrictive given its crystal structure. This implies that the strength of an HCP rubble pile against a tidal encounter increases with the number of particles composing it, in contrast to the previous HSDEM results of \cite{Richardson2005}. Since the friction is not physically modeled in the HSDEM, the rubble pile can more easily overcome the cannonball stacking and start to flow when $N$ becomes large enough. However, in the SSDEM, the interparticle friction is independent of the particle resolution, and the same frictional resistance is exerted on every particle pairs to hinder their relative movement. For the HCP configuration, as each particle band contains more particles with a higher resolution, the cumulative resistance acting on the whole band is larger, leading to a stronger structure. 
 
\subsection{Spin-shape evolution and comparison with continuum theory}
\label{s:results:continuum}

  \begin{figure}
    \centering
    \includegraphics[width = 8.8 cm]{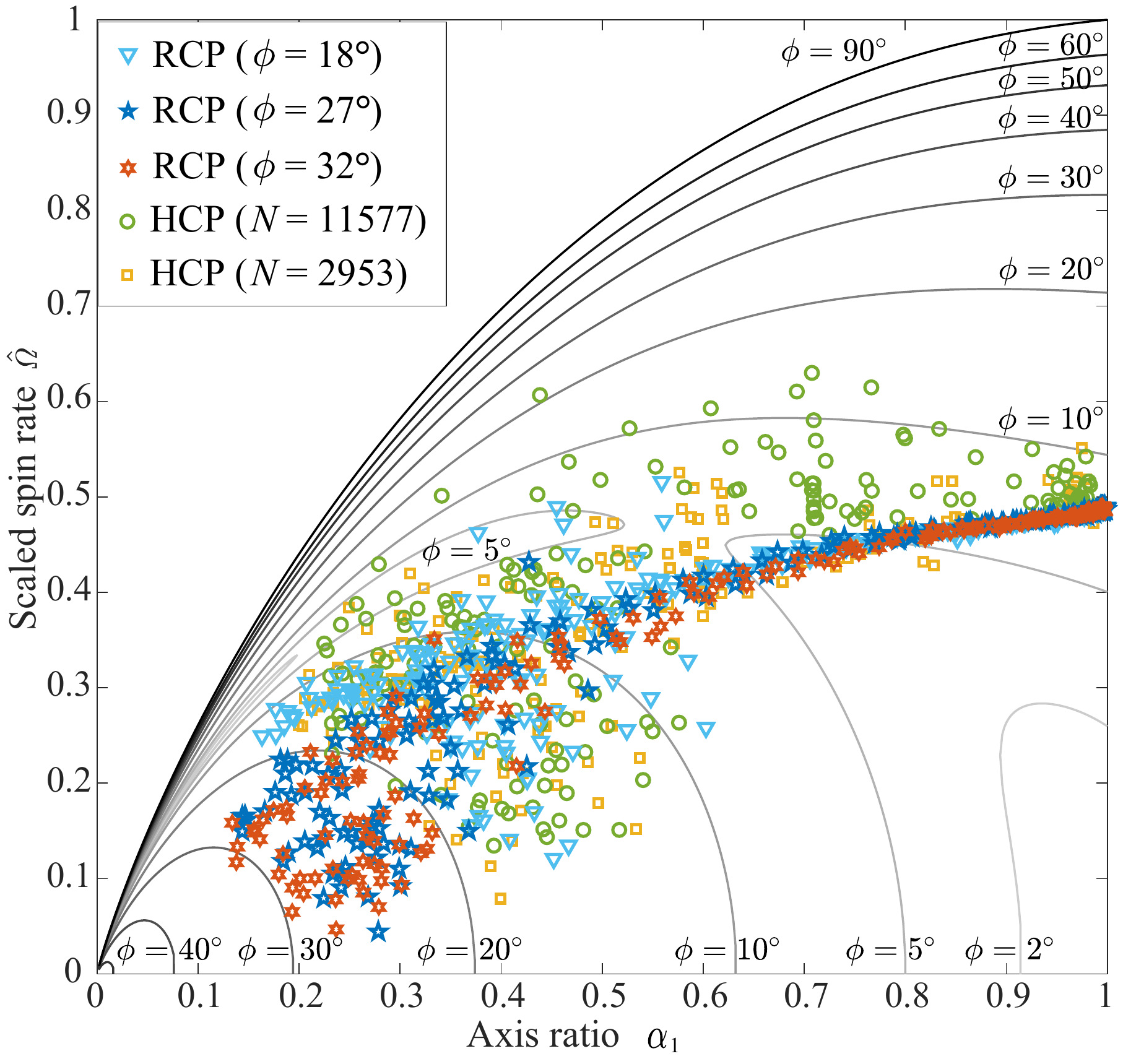}
    \caption{Scaled spin limits for prolate bodies. The colored symbols are the final stable spin rates of the largest remnants in our tidal encounter simulations using an RCP configuration with three different angles of friction, as well as an HCP configuration with two different resolutions, as a function of the largest remnants' axis ratios $\alpha_1$. The gray curves are the theoretical spin limits for the indicated angles of friction derived from a static continuum approach using the Drucker–Prager yield criterion \citep{Holsapple2007}. For angles of friction with two curves, the permissible spin rates lie in the region between the two curves. }
    \label{f:lr_distribution}
   \end{figure}

  \begin{figure}
    \centering
    \includegraphics[width = 8.8 cm]{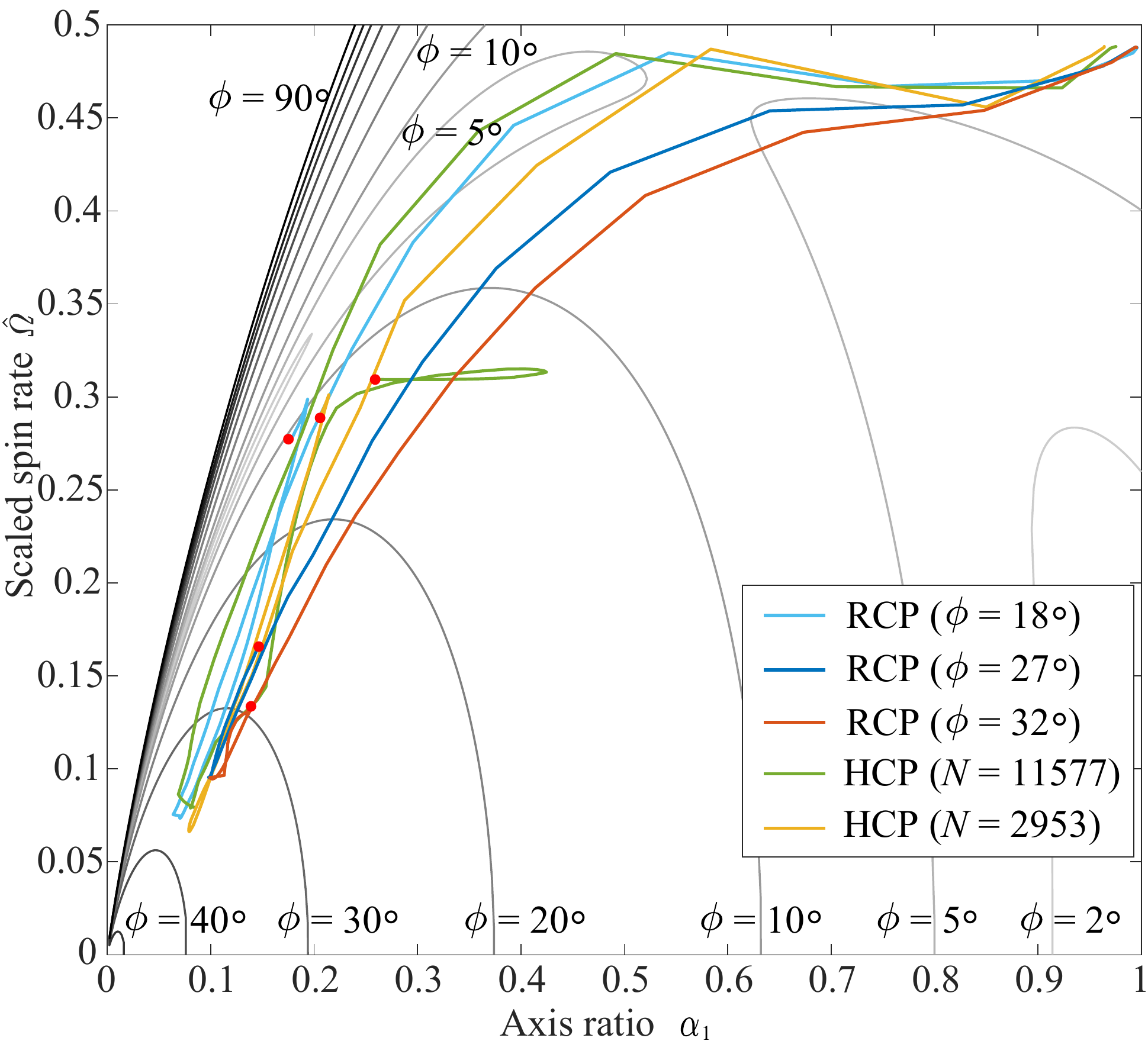}
    \caption{Dynamical evolution of the scaled spin rate during a tidal encounter with $V_{\infty}=11$ km/s and $q=1.2R_\oplus$. The colored curves show the dynamical evolution of the scaled spin rates of the largest remnants for the RCP models with three angles of friction and the HCP models with two resolutions, as a function of the rubble pile axis ratio $\alpha_1$. The red dot denotes the final stable state of the largest remnant for every case. The gray curves have the same meanings as in Fig.~\ref{f:lr_distribution}. }
    \label{f:spin_path}
   \end{figure}

 The permissible shape and spin limit of a rubble pile depend on its material strength. Based on static elastic-plastic continuum theory for solid materials, analytic expressions for the permissible spin rates as functions of the friction angle, $\phi$, for ellipsoidal bodies are presented in the work of \cite{Holsapple2004, Holsapple2007}. As shown in Fig.~\ref{f:lr_distribution}, for each friction angle, this theory gives two curves, one upper limit and one lower limit, and all spin-shape states in between are possible equilibrium states.
 
 We compare the final stable spin-shape states of the largest remnants obtained by our simulations with this theory. In Fig.~\ref{f:lr_distribution}, the scaled spin rate, expressed as $\hat{\Omega}= \omega / \sqrt{4 \pi \rho G/3}$, is represented as a function of the axis ratio in a diagram that allows direct comparisons with the continuum theory. $\omega$ and $\rho$ denote a body's spin rate and bulk density, respectively, and $G$ is the gravitational constant. As most of the largest remnants have prolate shapes, the theoretical spin limits are derived for prolate bodies whose axis ratios $\alpha_1=\alpha_2$. For the RCP models, it is clear to see that for a given angle of friction, the final spin states of the largest remnants always lie within the limits given by the continuum theory. For the HCP model, the initial angle of friction is $\sim$40$^\circ$. In some ``deformation'' events, parts of the internal interlocking features are preserved and can keep the largest remnants stable at states beyond the 18$^\circ$ limits. In ``shedding'' and ``disruption'' events, the tidal encounter breaks the HCP configuration and turns it into an RCP one with an angle of friction of about $\sim$18$^\circ$. Therefore, their largest remnants stay within the limits imposed by the continuum theory for this angle. 

 One big advantage of numerical simulations against the continuum theory is that we can study the spin-shape evolutionary dynamics of rubble-pile bodies in the tidal encounter process. Figure \ref{f:spin_path} shows the dynamical evolution of the clump of ``bound particles'' that will eventually form the largest remnant for different rubble-pile models for a given encounter with $V_\infty=11$ km/s and $q=1.2R_\oplus$ (which causes a ``shedding'' event with a few particles loss). For each case, the evolution starts from the top right of the plot of Fig.~\ref{f:spin_path} and ends at the red dot. Figure \ref{f:spin_path} can be analyzed in association with Figs.~\ref{f:velocityevolution}, \ref{f:tidalprocess_vinf5}, and \ref{f:tidalprocess_q16}. In the beginning of the tidal encounter, the tidal forces induce the outward movement of particles and cause the rapid elongation of the rubble pile (see Fig.~\ref{f:velocityevolution}b1, b2). At the same time, the tidal torques also act on it to maintain its rotation. Therefore, the spin rate of the clump remains similar for some time. Since the low-friction rubble-pile models are more deformable in this stage, their axis ratios appear to decrease more quickly. Then, as the distance to the planet increases, the tidal effects become inefficient and the clump only evolves under its self-gravity. Because the velocities of particles at the edge remain directed outward for some time (see Fig.~\ref{f:velocityevolution}c1, c2), the clump keeps elongating and therefore its spin rate slows down. When the spin rate is slow enough, the particles composing the clump reaccumulate to form a less elongated rubble pile, which leads to an increase of the spin rate until the rubble pile reaches an equilibrium configuration (see Fig.~\ref{f:velocityevolution}d1, d2). For a high angle of friction (e.g., the red curve in Fig.~\ref{f:spin_path}), the equilibrium is quickly established, while for a low angle of friction (e.g., the light blue curve in Fig.~\ref{f:spin_path}), the particles can flow more efficiently and need more time to damp out the excessive energy during the reaccumulation process, which results in a larger shift in the spin-shape diagram until equilibrium.

\section{Comparison between HSDEM and SSDEM}
\label{s:ComparingHS}

  \begin{figure}
    \centering
    \includegraphics[width = 8.85 cm]{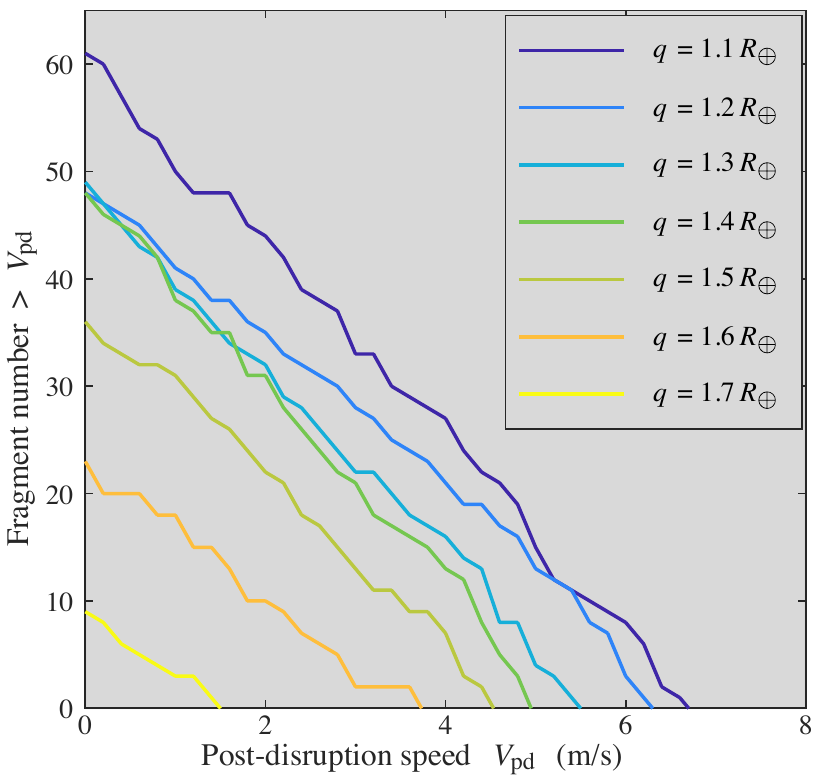}
    \caption{Cumulative fragment post-disruption speed distributions for the low-resolution HCP model in ``disruption'' events with $V_\infty=1$ km/s and different $q$ at the end of the SSDEM simulations (well beyond Earth’s Roche limit). The colors of the curves represent the results with different $q$ as indicated in the legend. The post-disruption speed is calculated for each fragment with respect to the largest remnant in each case. }
    \label{f:pd_speed}
   \end{figure}
   
  \begin{figure}
    \centering
    \includegraphics[width = 8.85 cm]{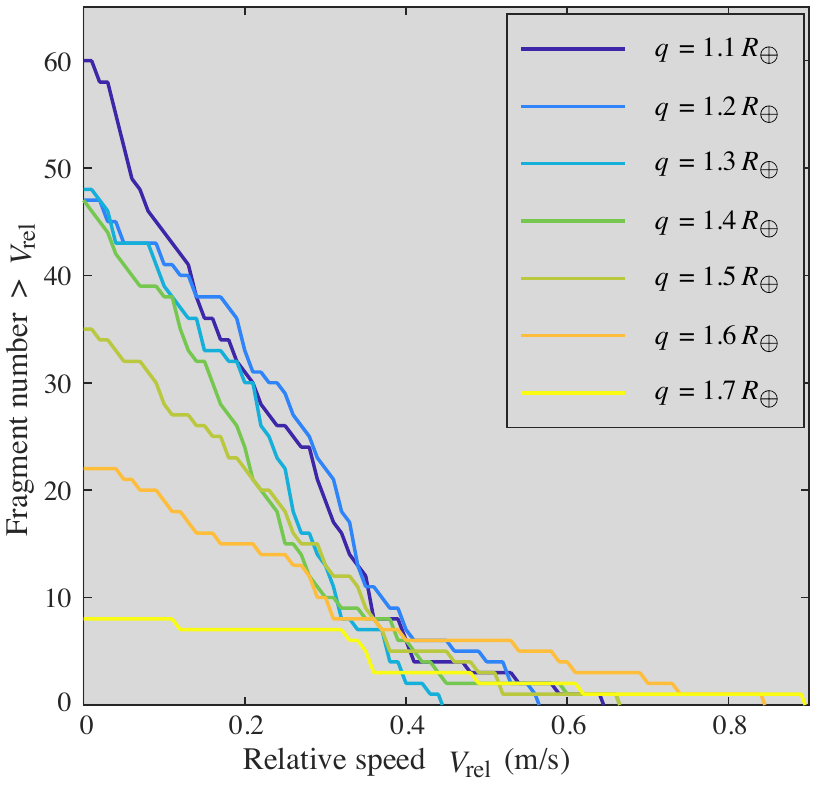}
    \caption{Cumulative fragment relative speed distributions for the low-resolution HCP model in ``disruption'' events with $V_\infty=1$ km/s and different $q$ at the end of the SSDEM simulations (well beyond Earth’s Roche limit). The colors of the curves represent the results with different $q$. The relative speed is calculated for each fragment with respect to its neighbors in the fragment trains for each case. }
    \label{f:rel_speed}
   \end{figure}
   
 Qualitatively, our SSDEM simulation findings on the behaviors (e.g., deformation and disruption) of rubble-pile bodies subject to tidal effects are similar to previous work using the HSDEM \citep[e.g.,][]{Richardson1998, Walsh2006, Schunova2014}. However, some detailed processes and many quantitative characteristics are very different between the results of these two methods. In Sect.~\ref{s:results:resolution}, we have shown some divergences about the effect of particle resolution between the simulation results using the HSDEM and the SSDEM. Here, we present a comprehensive comparison between these two methods in terms of the tidal encounter outcomes.  Since we intentionally used the same setup as the HSDEM tidal encounter simulations of \citet{Schunova2014} for a one-to-one comparison (see Sect.~\ref{s:method}), their simulation results are used as a main source for HSDEM results. Our simulation results of the low-resolution HCP model, which is analogous to their rubble-pile progenitor models (see Sect.~\ref{s:method:rubblepile}), are used as comparisons. 

\subsection{Mass loss and tidal disruption frequency}

 As introduced in Sect.~\ref{s:introduction}, the study by \citet{Schunova2014} was aimed at estimating the properties and evolution of NEO families that originate from the tidal disruption of a NEO progenitor by the Earth. By running tidal encounter simulations with an HCP rubble-pile model on 718 different Earth encountering orbits with $V_\infty\lesssim35$ km/s and $q\lesssim2.5R_\oplus$, they obtained the encounter conditions that can induce different levels of mass loss. 
 
 According to Fig.~1 of \citet{Schunova2014}, the HSDEM tidal-induced mass loss as a function of $q$ and $V_\infty$ is presented in the top-left panel of Fig.~\ref{f:resolutioneffect}. Although the mass-loss trend in terms of $q$ and $V_\infty$ is similar, that is, more mass is lost for a smaller $q$ or $V_\infty$, HSDEM simulations drastically overestimate the efficiency of Earth close encounters in causing mass loss of the progenitor. For instance, in SSDEM simulations, for $V_\infty=0$ km/s, a rubble pile starts losing mass at a periapse distance $q\sim1.9R_\oplus$, while mass loss occurs at periapse distances much greater than $2.5R_\oplus$ in HSDEM simulations. In general, compared with the HSDEM simulations, all the mass loss results of the SSDEM simulations shift towards much lower periapse distances in this plot. This means that the frequency of NEO families resulting from tidal disruption by the Earth should be much lower that estimated by the results based on the HSDEM \citep[e.g., ][derived the frequency of tidal disruption for NEOs to be about once every $10^5$ years based on their HSDEM simulations]{Richardson1998}, and should be reassessed based on SSDEM simulation results. 

\subsection{Speed distribution of fragments}

 In tidal encounter ``disruption'' events, the speed distribution of the resulting fragments determines their relative motion and, thus, it is important for predicting their dynamical evolution and formation of multiple-body systems.
 
 In order to access the post-disruption orbital evolution of tidal-induced NEO families, \citet{Schunova2014} analyzed the post-disruption speed of members of the NEO families created by their HSDEM tidal disruption simulations. The maximum post-disruption speed with respect to the largest remnant in each family is about 4 m/s (see their Fig.~3). The magnitude of the post-disruption speed is important for them to derive their main findings, for example, the detectable lifetime of the tidal-induced NEO families. Figure~\ref{f:pd_speed} presents the cumulative post-disruption speed distributions of the resulting fragments in our SSDEM simulations with $V_\infty=1$ km/s. The post-disruption speed $V_\mathrm{pd}$ is proportional to the distance to the largest remnant and, therefore, the distribution is rather uniform. As the progenitor is subject to more severe tidal disruption with a smaller $q$, the maximum post-disruption speed increases rapidly with a closer encounter. The most intensive encounter conditions shown in \citet{Schunova2014} are similar to the case of $V_\infty=1$ km/s and $q=1.5R_\oplus$, which generates a similar maximum post-disruption speed $\sim$4 m/s. However, our results imply that a closer or farther approach would significantly increase or decrease the post-disruption speed, leading to a shorter or longer NEO-family detectable lifetime. These effects have not been considered in \citet{Schunova2014}. Further studies based on the SSDEM are important for revealing the dynamical evolution of tidal-induced NEO families. 

 We also analyzed the relative speed of the fragments with respect to their neighbors in the resulting fragment trains. As shown in Fig.~\ref{f:rel_speed}, since the trains contain fewer fragments, the maximum relative speed increases with a closer encounter. The relative speeds of some fragment pairs are smaller than 0.1 m/s, which is less than the escape speed of a hundred-meter-sized body. These fragments could potentially merge into bifurcated bodies or form binary asteroid systems. \citet{Schunova2014} reported that their HSDEM tidal disruption simulations generate an averaged relative speed between fragments of about 0.4 m/s at the initial stage and this value increase to about 0.6 m/s at the end of simulations. This averaged relative speed between the fragments is notably larger than the results obtained by our SSDEM simulations, which is about 0.2 m/s. This implies that a small amount of interparticle friction and the proper treatment of physical contact processes indeed can hinder the relative movement between fragments. The binary formation frequency may be much higher than previous predictions using the HSDEM \citep[e.g.,][]{Walsh2006, Walsh2008}, and the physical and orbital properties of the resulting binaries may be closer to some binary NEO systems than the HSDEM predicted. Further studies quantifying tidal encounter outcomes could significantly improve our understanding of the NEO evolution and population as a whole.

\subsection{Shapes of fragments}
 
 Previous works using the HSDEM have successfully reproduced the shapes of some elongated and bifurcated asteroids through a tidal encounter \citep[e.g., 1620 Geographos and 433 Eros by][]{Solem1996, Bottke1999}. The maximum elongated remnant obtained by the HSDEM reported in the literature \citep[see Fig.~9 in ][]{Walsh2006} has an elongation of $\sim$0.72, close to our maximum elongated remnant of $\varepsilon_\mathrm{lr}\sim0.78$ for a similar progenitor model (i.e., the low-resolution HCP model; see the top-middle panel of Fig.~\ref{f:resolutioneffect}). Compared to the low production rate of such extremely elongated remnants by the HSDEM (lower than 1\%), the SSDEM is more robust in producing highly elongated shape. With higher friction angles, the SSDEM is even able to generate rubble piles with elongations $\sim$0.85 (e.g.,  the bottom-middle panel of Fig.~\ref{f:frictioneffect}). Furthermore, since the SSDEM is better at maintaining the structural stability of an irregular shape \citep{Zhang2017, Hu2018}, the tidal effects on the detailed morphology of small bodies can be revealed \citep{DeMartini2019}. Further studies to characterize the shape in tidal disruption outcomes could reveal the evolutionary history of some NEOs.

\section{Application to SL9 and comparison with a polyhedral method}
\label{s:sl9}

  \begin{figure*}
    \centering
    \includegraphics[width = 18 cm]{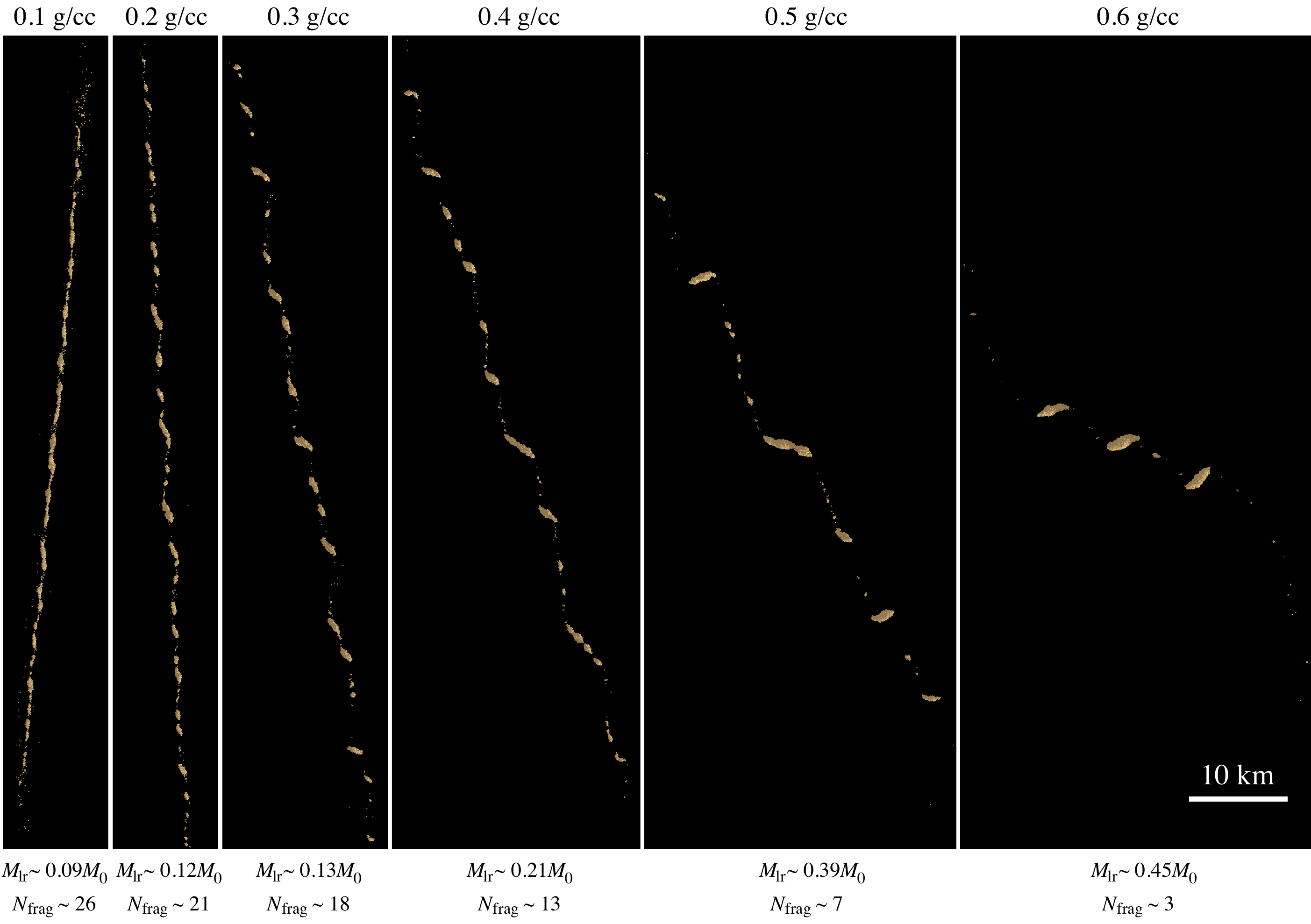}
    \caption{Snapshots of fragment distributions in the SL9 tidal encounter simulations with different progenitor bulk densities (as indicated on the top of each snapshot), shown $\sim$14 hours after perijove (about to pass the orbit of Ganymede). The mass of the largest remnant $M_\mathrm{lr}$ and the number of fragments with mass larger than 1\% of the original mass $M_0$ are given on the bottom for each density case.}
    \label{f:sl9}
  \end{figure*}
  
  In this section, we apply our SSDEM to study the most famous tidal disruption event, the encounter of comet Shoemaker-Levy 9 (SL9) with Jupiter in 1992. This event has provided valuable information to benchmark the physical properties of small bodies.  \cite{Asphaug1994,Asphaug1996} performed the numerical modeling of this disruptive encounter using two kinds of approaches and progenitor models: a non-rotating monolithic sphere modeled with their SPH code with tensile strength \citep{Benz1994}, and a spherical rubble-pile aggregate consisting of frictionless equal-sized spheres modeled with an $N$-body code with soft elastic collisions. A better match was found with the rubble-pile SL9 model and the estimated bulk density for SL9 is about 0.6 g/cc according to the resulting chain morphology. A more recent study by \cite{Movshovitz2012}, using polyhedra instead of spheres to represent the components of rubble-pile SL9, found that the inclusion of irregular shapes for the constituent particles enhances relative rotational resistance and the whole body can experience dilatancy. Consequently, the body is stronger against tidal forces than a rubble pile made with spheres. As a result, a much lower bulk density (0.3--0.4 g/cc) is suggested for SL9.

  The parameter $\beta$ and the associated spring-dashpot-slider rotational resistance model in our SSDEM allow capturing some aspects of the behavior of a rubble pile made of components with irregular shapes \citep{Zhang2017}. As shown in Sect.~\ref{s:results:material_effect}, increasing the value of $\beta$ can lead to larger fragments in moderate disruption encounters. This factor has not been considered in previous sphere-based methods, such as the one used in \cite{Asphaug1996} and \cite{Movshovitz2012}, to simulate tidal encounters. 
  
  To compare the behavior of a rubble-pile model made of polyhedra and our rubble-pile model made with spheres, including this rotational resistance prescription, we performed simulations of the SL9 tidal encounter, for different assumed bulk densities and compared them with the results of \cite{Movshovitz2012}. To make a one-to-one comparison, we use the same encounter conditions and a similar rubble-pile model as those used in \cite{Movshovitz2012}. In particular, SL9 is modeled as a non-rotating, 1-km-sized, random-close-packing rubble pile consisting of $\sim$4,000 identical spheres. The encounter orbit is nearly parabolic with eccentricity of 0.997 and perijove distance of 1.33 Jovian radii. To mimic the effect of irregular particle shapes, we adopted the high shear strength parameter set ($\mu_S=1.0$, $\beta=0.8$). The corresponding friction angle $\phi=32^\circ$ is common for sands and is smaller than local slopes measured for some asteroids visited by spacecraft \citep[e.g., the surface slope of asteroid Bennu exceeds $40^\circ$ in some regions; see][]{Barnouin2019}.
  
  Figure \ref{f:sl9} shows the outcomes of our simulations, $\sim$14 hours after passage at perijove. For a low bulk density, the disruption leads to many small fragments and a string or chain-like pattern. As the bulk density of the progenitor increases, the number of fragments formed by the tidal disruption decreases and the mass of the largest remnant increases. The fragment distributions and the dependency on the density are qualitatively consistent with the results of \cite{Movshovitz2012} (e.g., see their Fig. 7), showing that our spherical model is capable of capturing the similar kind of dynamics as a polyhedral one. We note, however, that the time after perijove at which we find such results is later ($\approx 14$ hr) than that in \cite{Movshovitz2012} ($\approx 5$ hr), as earlier, the disruption process of our rubble pile is not yet complete. This indicates that the post-disruption speed is lower in our case. With the appropriate consideration of contact physics and dynamics in a granular material, our SSDEM can help to delay breakup and slow down the relative motion between particles. Moreover, the bulk density for which the behavior of our rubble pile looks similar to the resulting chain morphology of SL9 is around 0.2--0.3 g/cc, which is lower than previous estimates using spheres but similar to the one using polyhedra. The consistency between our results and the results using the polyhedral approach validates the capability of our SSDEM in capturing the contact dynamics of irregular shapes of realistic particles. 
  
  We further ran extra simulations with different material parameters $\mu_S$ and $\beta$ for the case with a bulk density of $0.3$ g/cc, and found that these two parameters have little impact on the fragment size distribution in such catastrophic tidal disruption events, as pointed out in Sect.~\ref{s:results:material_effect}. This finding agrees with the work of \cite{Movshovitz2012}, who found that the friction coefficient in their model does not strongly affect the outcome of an “SL9-type” tidal encounter. 
  
\section{Conclusions and perspectives}
\label{s:conclusions}
 
 We performed new simulations of tidal encounters of rubble-pile progenitors with a planet, using the soft-sphere discrete element method (SSDEM). We considered two different packing, that is, a hexagonal close packing (HCP) and a random close packing (RCP), to model the rubble-pile progenitors. The effect of the particle resolution and the material strength on the encounter outcomes were explored.

 Our results are very different from the results of earlier studies using the hard-sphere discrete element method \citep[HSDEM; e.g., ][]{Richardson1998, Walsh2006, Schunova2014}. In particular, we found that the latter method overestimates the efficiency of planetary approaches to cause a mass loss of the rubble pile undergoing tidal forces. Therefore, asteroid family formation resulting from tidal disruption cannot be as frequent as previously estimated using the HSDEM. The shapes of resulting fragments are much more elongated and their post-encounter evolution indicates higher production rates for bifurcated bodies and binaries according to our SSDEM results. Our study suggests that there are many observed characteristics of small bodies that can be linked to tidal effects. 
 
 Furthermore, the comparison of our simulations of SL9 tidal disruption with a previous numerical study using polyhedra rubble-pile models \citep{Movshovitz2012} shows a good match. We thus conclude that appropriately accounting  for the contact dynamics and the rotational resistance in rubble-pile modeling with spherical particles allows us to capture the dynamics of rubble piles made of components with irregular shapes. As the computational efforts using spheres-based models are much less stringent than those using polyhedra-based models, our SSDEM provides a very efficient alternative to model rubble-pile physics without losing important features. It is nevertheless possible that the use of spherical shapes, even considering high friction, may not capture the influence of irregular shapes in some particular conditions or contexts. Thus, other studies using other implementations with irregular shapes would be useful to check against when such prescriptions are necessary. For instance, in a future study, we would use our SSDEM code and replace spherical particles with rigid aggregates made of such particles in our SSDEM code, as done in \cite{Richardson2009} and \cite{Thuillet2019}, to investigate under which conditions, and in which contexts, modeling explicitly irregular shapes is required. 
 
 We also found that the particle configuration in a rubble pile has great influence on the encounter outcomes. With the same encounter conditions, the RCP model generally leads to more severe changes due to tidal forces than the HCP model. In response to tidal forces, the RCP model is shown to behave in a sensitive fashion with regard to its material shear strength and demonstrates a continuous trend with decreasing periapse distance and speed at infinity, while a tidal encounter of the HCP model strongly depends on the precise internal configuration and orientation towards the tidal force directions.

 The strength of an HCP progenitor against a tidal encounter also depends  on the number of particles composing it, while the effect is less pronounced for a random packing rubble-pile progenitor. For the HCP model, we found that fine configurations consisting of a larger number of particles are more resistant to tidal disruption than coarse configurations with fewer particles, in contrast to the previous HSDEM results of \cite{Richardson2005}, as the cumulative friction in an HCP structure is increased with the particle number in the SSDEM.

 In general, we find that tidal encounters do not lead to any apparent reshaping or disruption for distances above the theoretical tidal failure limiting distance predicted by \cite{Holsapple2006}. Our simulation outcomes also find a good agreement with the permissible shape and spin limit of rubble piles given by the continuum theory of \cite{Holsapple2004, Holsapple2007}. Through the study of the spin-shape evolutionary paths, we revealed the detailed dynamical response of rubble-pile bodies to tidal encounters and the mechanisms leading to extremely elongated shapes of small bodies.

 Our investigation shows that predicting what will happen during the tidal encounter of a small body requires some knowledge about the internal configuration of the object.  We thus encourage further studies aimed at exploring the tidal processes and outcomes using the SSDEM that would also continue to explore a wide parameter space. Furthermore, measuring the end result of a tidal encounter can allow us to provide some constraints on the internal and mechanical properties of the object, as has been done for SL9. Monitoring a very close planetary approach, such as the asteroid Apophis encounter with the Earth in 2029, could provide clues about the mechanical and internal structures of this asteroid, which would give insights into this gold mine of information that would benefit our understanding of how these bodies form and later evolve in response to external processes.

\begin{acknowledgements}
      We thank Kevin Walsh and Derek C. Richardson for inspiration and suggestions on exploring the differences between HSDEM and SSDEM for simulating close encounters. We also thank the anonymous referee for the constructive reviews. Y. Z. acknowledges funding from the Universit\'e C\^ote d'Azur ``Individual grants for young researchers'' program of IDEX JEDI.  Y. Z. and P. M. acknowledges support from CNES and from the Acad. 2 and 3 of Univ. Côte d’Azur IDEX JEDI. This project has received funding from the European Union’s Horizon 2020 research and innovation program under grant agreement No. 870377 (project NEO-MAPP). Simulations were performed on the LICALLO cluster hosted at the Observatoire de la C\^ote d'Azur and on the yorp cluster administered by the Department of Astronomy and the Deepthought2 HPC cluster administered by the Division of Informational Technology at the University of Maryland.  
\end{acknowledgements}

\bibliographystyle{aa} 
\bibliography{references.bib} 

\end{document}